\documentclass[fleqn,usenatbib]{mnras}

\usepackage{newtxtext,newtxmath}

\usepackage[T1]{fontenc}
\usepackage{ae,aecompl}

\usepackage{graphicx}	
\usepackage{amsmath}	

\usepackage{multirow}
\usepackage{adjustbox}

\usepackage[verbose]{placeins}
\usepackage{blindtext}

\usepackage{bm}
\usepackage{mathtools}
\usepackage{xcolor}
\usepackage{natbib}
\usepackage{lscape}
\usepackage{afterpage}  
\usepackage{booktabs}
\usepackage{tabularx}
\usepackage{threeparttable}

\usepackage{newtxtext,newtxmath}

\title[A strong LyC leaker at $z=3.24$]{The UV-brightest Lyman continuum emitting star-forming galaxy}

\author[R. Marques-Chaves et al.]{
R. Marques-Chaves$^{1}$\thanks{E-mail: Rui.MarquesCoelhoChaves@unige.ch},
D. Schaerer$^{1,2}$, 
J. \'{A}lvarez-M\'{a}rquez$^{3}$,
L. Colina$^{3,4}$, 
\newauthor 
M. Dessauges-Zavadsky$^{1}$,
I. P\'{e}rez-Fournon$^{5,6}$,
A. Saldana-Lopez$^{1}$,
A. Verhamme$^{1}$%
\\
$^{1}$Geneva Observatory, University of Geneva, Chemin Pegasi 51, CH-1290 Versoix, Switzerland\\
$^{2}$CNRS, IRAP, 14 Avenue E. Belin, 31400 Toulouse, France\\
$^{3}$Centro de Astrobiolog\'ia (CSIC-INTA), Carretera de Ajalvir, 28850 Torrej\'on de Ardoz, Madrid, Spain\\
$^{4}$International Associate, Cosmic Dawn Center (DAWN) \\
$^{5}$Instituto de Astrof\'\i sica de Canarias, C/V\'\i a L\'actea, s/n, E-38205 San Crist\'obal de La Laguna, Tenerife, Spain\\
$^{6}$Universidad de La Laguna, Dpto. Astrof\'\i sica, E-38206 San Crist\'obal de La Laguna, Tenerife, Spain\\
}

\date{}

\begin{document}
\label{firstpage}
\pagerange{\pageref{firstpage}--\pageref{lastpage}}
\maketitle

\begin{abstract}
We report the discovery of J0121$+$0025, an extremely luminous and young star-forming galaxy ($M_{\rm UV} = -24.11$, log[$L_{\rm Ly \alpha} / \rm erg~s^{-1}] = 43.8$) at $z=3.244$ showing copious Lyman continuum (LyC) leakage ($f_{\rm esc, abs} \approx 40\%$).
High signal-to-noise ratio rest-frame UV spectroscopy with the Gran Telescopio Canarias reveals a high significance ($7.9\sigma$) emission below the Lyman limit ($< 912$\AA), with a flux density level $f_{900} = 0.78 \pm 0.10 \mu$Jy, and strong P-Cygni in wind lines of O~{\sc vi}~1033\AA, N~{\sc v}~1240\AA{ }and C~{\sc iv}~1550\AA{ }that are indicative of a young age of the starburst ($<10$~Myr). 
The spectrum is rich in stellar photospheric features, for which a significant contribution of an AGN at these wavelengths is ruled out. Low-ionization ISM absorption lines are also detected, but are weak ($EW_{0} \rm  \simeq 1$\AA) and show large residual intensities, suggesting a clumpy geometry of the gas with a non-unity covering fraction or a highly ionized ISM. The contribution of a foreground and AGN contamination to the LyC signal is unlikely. Deep optical to Spitzer/IRAC $4.5\mu$m imaging show that the spectral energy distribution of J0121$+$0025 is dominated by the emission of the young starburst, with log($M_{\star}^{\rm burst}/M_{\odot}) = 9.9\pm 0.1$ and $\rm SFR = 981\pm232$~$M_{\odot}$~yr$^{-1}$. J0121$+$0025 is the most powerful LyC emitters known among the star-forming galaxy population. The discovery of such luminous and young starburst leaking LyC radiation suggests that a significant fraction of LyC photons can escape in sources with a wide range of UV luminosities and are not restricted to the faintest ones as previously thought. These findings might shed further light on the role of luminous starbursts to the cosmic reionization.
\end{abstract}




\section{Introduction}

Lyman-$\alpha$ emitters (LAEs) and Lyman break galaxies (LBGs) are 
widely studied populations of star-forming galaxies that are common in the early Universe. Deep field surveys have been used to study these star-forming galaxies, which are typically faint with magnitudes $R \sim 25$~AB at $z \sim 3$, corresponding to absolute UV magnitudes $M_{\rm UV}^{*}$ of about $-20$ to $-21$ at those redshifts \citep[for LAEs and LBGs, respectively; e.g.,][]{ouchi2008, reddy2009}.
These studies have revealed that typical ($M_{\rm UV}^{*}$) LAEs and LBGs present a wide range of properties, with stellar masses $\log(M_{\star}/M_{\odot}) \sim 8.0-10.0$ \citep[e.g.,][]{shapley2001, gawiser2007, ono2010a, santos2019}, star-formation rates (SFR) up to a few tens $M_{\odot}$~yr$^{-1}$ \citep[e.g.,][]{shapley2003, nakajima2012, sobral2018b}, and low metallicity on average \citep[e.g.,][]{finkelstein2011a, nakajima2013, kojima2017}. 

These star-forming galaxies are likely the dominant sources responsible for ionizing the intergalactic medium (IGM) in the early Universe, during the so-called Epoch of Reionization (EoR at $6 < z < 15$; e.g., \citealt{robertson2015}), due to their expected large hydrogen ionizing photon (hereafter Lyman continuum; LyC with $>13.6$ eV) escape fraction, $f_{\rm esc} \rm (LyC)$. However, it is still unclear whether the faint and numerous, or the more luminous and rare are the main contributors \citep[cf.,][]{finkelstein2019, Naidu2020}. 

Recent progress has been made in selecting LyC emitters and understanding their properties. 
To date $\approx 50$ LyC emitting star-forming galaxies have been discovered and studied in detailed at low-$z$ ($z \lesssim 0.4$; e.g., \citealt{borthakur2014, izotov2016, leitherer2016, izotov2018a, izotov2018b}, intermediate-$z$ ($z\sim 1.4$, \citealt{saha2020}, and moderately high-$z$ ($z>2$, \citealt{debarros2016, shapley2016, vanzella2016, bian2017, steidel2018, fletcher2019, rivera2019,  ji2020}) up to $z\sim 4$ \citep{vanzella2018}, where the IGM still allows the direct observation of LyC photons. However, the relation between $f_{\rm esc} \rm (LyC)$ and $M_{\rm UV}$ is still a matter of debate. Some statistical studies targeting star-forming galaxies with $M_{\rm UV}$ between $\simeq -20$ to $-21.5$ find a clear trend, where faint galaxies show larger $f_{\rm esc} \rm (LyC)$ than luminous ones \citep{steidel2018, pahl2021}. However, these findings are in tension with those from other works. For example, \cite{bian2020} find a low $f_{\rm esc} \rm (LyC) < 14\%$ ($3\sigma$) in faint LAEs ($M_{\rm UV} \simeq -18.8$), suggesting that the LyC leakage in star-forming galaxies do not follow such trend. In addition, one of the strongest LyC emitters known at high redshift, with $f_{\rm esc} \rm (LyC) \approx 60\%$ is also a very luminous ($M_{\rm UV} = -22.20$) star-forming galaxy (\textit{Ion3} at $z = 4.0$; \citealt{vanzella2018}).

Understanding the role of UV-luminous star-forming galaxies to the cosmic reionization requires the knowledge of two key properties: the volume density of these sources and their intrinsic properties, those governing the production  (e.g., stellar population, star-formation histories) and escape of LyC photons (i.e. neutral gas and dust content and the geometry of the interstellar medium, ISM). 
UV-luminous star-forming galaxies are rare \citep[e.g.,][]{sobral2018b}, yet how much is still unknown, in particular at $z>6$.
Recent studies have found remarkably bright galaxies well into the EoR ($z>7$) that are in excess compared to the generally observed Schechter component of the UV luminosity function (LF; e.g., \citealt{bowler2014, bowler2015, ono2018, stefanon2019}). For example, the most distant spectroscopically-confirmed galaxy known, \textit{GN-z11} at $z \simeq 11.0$ \citep{oesch2016, jiang2021}, is also very luminous in the UV ($M_{\rm UV} = -22.1$), and the inferred volume density of this source is a factor of $> 15$ higher than predicted by models \citep{oesch2016}.

On the other hand, the number of UV-luminous star-forming galaxies known is scarce, in particular those as bright as $M_{\rm UV} < -23$ \citep[e.g.,][]{des2010, lee2013b, harikane2020, marques2020}, preventing us to study their physical properties in a statistical basis. 
While bright sources are in principle easy and optimal targets for deep follow-up observations and detailed analysis of their properties, identifying them is extremely challenging, as it requires large area surveys and, in addition, spectroscopic follow-up to distinguish them from the much more abundant active galactic nuclei (AGN), which is time consuming.

Motivated by this, we have undertaken a search for UV-luminous star-forming galaxies at $z>2$ in of one of the widest spectroscopic surveys ever performed, the $\sim 9300$~deg$^{2}$-wide extended Baryon Oscillation Spectroscopic Survey \citep[eBOSS:][]{abolfathi2018} of the Sloan Digital Sky Survey \citep[SDSS:][]{eisenstein2011}. 
While eBOSS was specifically designed to select bright quasi stellar objects (QSOs) at these redshifts, not necessarily high-$z$ star-forming galaxies, some QSO candidates were selected and observed based on their optical colors only \cite[for details see:][]{ross2012}, mimicking the Lyman break technique. The combination of a wide area surveyed, the limiting $r$-band magnitude of $\sim 22$~AB, which corresponds to $M_{\rm UV} \lesssim -23$ at $z\simeq 2.5$, and the available optical spectra for every source, makes eBOSS an alternative and efficient survey to search and explore such luminous star-forming galaxies.  

In \cite{marques2020b} we presented the first result of this project, where an extremely luminous star-forming galaxy was analysed, BOSS-EVULG1 at $z=2.47$. First selected as a SDSS QSO, follow-up observations revealed that the large luminosities in the UV ($M_{\rm UV} = -24.4$) and nebular emission (log(L[Ly$\alpha$, erg~s$^{-1}$]) = $44.0$) in BOSS-EVULG1 are powered by a vigorous starburst with SFR $\simeq$ 1000~$M_{\odot}$~yr$^{-1}$, a young age ($\simeq 4$~Myr) and very high sSFR ($\sim 100$~Gyr$^{-1}$), with no evidence for a dominant contribution of a type-I/II AGN. Interestingly, BOSS-EVULG1 shows properties that are common in LyC leakers, such as low dust content ($E(B-V) \simeq 0.07$ and log($L_{\rm IR}$/$L_{\rm UV}) < -1.2$) and a highly ionized ISM ([O~{\sc iii}]~5007\AA~/~[O~{\sc ii}]~3728\AA~$ \simeq 18$). In addition, it presents very weak low-ionization ISM absorption, suggestive of low geometric covering fraction of the neutral gas, that could be related with the powerful starburst-driven ionized outflow detected in H$\alpha$ in this galaxy \citep{alvarez2021}. However, no direct information on the LyC leakage is known for this luminous source.

In this work, we present the discovery and detailed analysis of a new luminous source, SDSS J012156.09$+$002520.3 ($\alpha$,~$\delta$ [J2000] = 20.4837$^{\circ}$, 0.4223$^{\circ}$), hereafter J0121$+$0025, an extremely UV-luminous ($M_{\rm UV} = -24.1$) star-forming galaxy at $z=3.244$ showing copious LyC leakage ($f_{\rm esc, abs} \approx 40\%$). The paper is structured as follows. The discovery and follow-up observations are presented in Section \ref{discovery}. The analysis of the rest-frame UV spectroscopic observations and imaging data are presented in
Section \ref{results}. In Section \ref{discussion} we discuss the LyC properties of J0121$+$0025 and compare them with those from other LyC emitters, and, finally, in Section \ref{conclusion} we present the summary of our main findings.
Throughout this work, we assume a \cite{salpeter1955} initial mass function (IMF) and a concordance cosmology with $\Omega_{\rm m} = 0.274$, $\Omega_{\Lambda} = 0.726$, and $H_{0} = 70$ km s$^{-1}$ Mpc$^{-1}$. All magnitudes are given in the AB system.

\begin{figure*}
  \centering
  \includegraphics[width=0.98\textwidth]{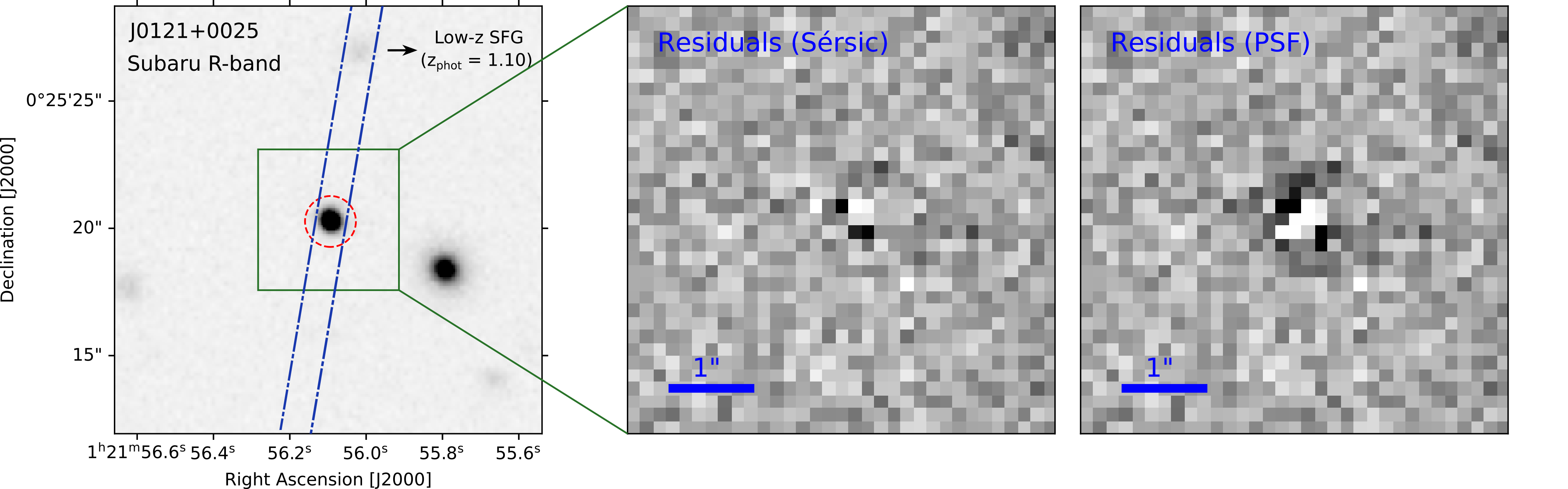}
  \caption{Cutout of J0121$+$0025 from the Subaru/HSC $R$-band image (left). The orientation of the GTC/OSIRIS long slit is marked in blue. The slit also includes a low-$z$ star-forming galaxy with a photometric redshift of $1.10\pm0.27$ and located $\simeq 6.6^{\prime \prime}$ North from J0121$+$0025. The spectroscopic $1^{\prime \prime}$-radius BOSS fiber is also marked in red. J0121$+$0025 shows a compact morphology ($\simeq 0.55^{\prime \prime}$~FWHM) and it is only barely resolved in this image (with a PSF of $\simeq 0.50^{\prime \prime}$~FWHM). Middle and right panels show the residuals from the Galfit modeling of J0121$+$0025 in the same band using a Sersic and PSF profiles, respectively. As shown in these panels, J0121$+$0025 is better modeled using a Sersic profile than a PSF, being the r.m.s of the residuals in the region encompassing J0121$+$0025 reduced by $\approx 50\%$.}
  \label{fig1}
\end{figure*}

\section{Discovery and Follow-up Observations}\label{discovery}

J0121$+$0025 was discovered as part of our search for Extremely UV-Luminous Galaxies (EUVLGs, $M_{\rm UV}$<$-$23) within the eBOSS survey \citep[][]{abolfathi2018} of the SDSS \citep[][]{eisenstein2011}. J0121$+$0025 is part of a large sample of $\sim$70 very luminous star-forming galaxies that were previously classified as QSOs in the Data Release 14 Quasar catalog \citep[][]{paris2018}. This sample also includes BOSS-EUVLG1 at $z=2.469$ recently analysed in \cite{marques2020b} and in \cite{alvarez2021}. The sample and the selection techniques will be presented in a separated work (R.~Marques-Chaves in prep.). Briefly, the selection consists on searching for narrow Ly$\alpha$ profiles in optical SDSS spectra of $z\gtrsim 2$ sources, and blue/flat optical to mid-IR colors, i.e., properties that are not expected to be present in QSOs. 

The BOSS spectrum of J0121$+$0025 (plate-mjd-fiberid: 4228-55484-818) shows features characteristic of an un-obscured, luminous star-forming galaxy, rather than an AGN. In particular, the shallow BOSS spectrum shows a narrow Ly$\alpha$ profile ($\simeq 350$~km~s$^{-1}$ full width half maximum, FWHM) and evidence of wind profiles in N~{\sc v} 1240\AA{ }and C~{\sc iv} 1550\AA{ }in the form of P-Cygni, that could be indicative of a young starburst.
In addition, the BOSS spectrum shows emission below $\lambda_{\rm obs} < 3870$\AA{ }, that although detected with low significance ($\simeq 2.3 \sigma$), could be an indication of LyC leakage ($\lambda_{\rm 0} < 912$\AA). Given this, J0121$+$0025 was selected as a priority target for deep follow-up spectroscopy.

\subsection{GTC/OSIRIS rest-frame UV spectroscopy}

Optical spectroscopy was obtained with the Optical System for Imaging and low-Intermediate-Resolution Integrated Spectroscopy instrument (OSIRIS)\footnote{\url{http://www.gtc.iac.es/instruments/osiris/}} on the 10.4~m Gran Telescopio Canarias (GTC). The data were obtained in service mode over two nights, 2020 August 18 and 19 in dark Moon and sub-arcsec ($\simeq 0.8^{\prime \prime}-0.9^{\prime \prime}$) conditions as part of the GTC program GTC21-20A (PI: R. Marques-Chaves). The R1000B grism was used, with dispersion of 2.12\AA, providing a full spectral coverage of 3600 - 7500\AA, which corresponds to 860 - 1800\AA{ }in the rest-frame at $z \simeq 3.25$. The OSIRIS 1.2$^{\prime \prime}$-wide slit was centered on J0121$+$0025 and oriented with the parallactic angle (see Figure \ref{fig1}). Given this configuration, the corresponding instrumental resolution is $\rm R\sim 800$ or $\simeq 400$~km~s$^{-1}$. In total 12 exposures of 900~s were acquired.

\begin{figure*}
  \centering
  \includegraphics[width=0.98\textwidth]{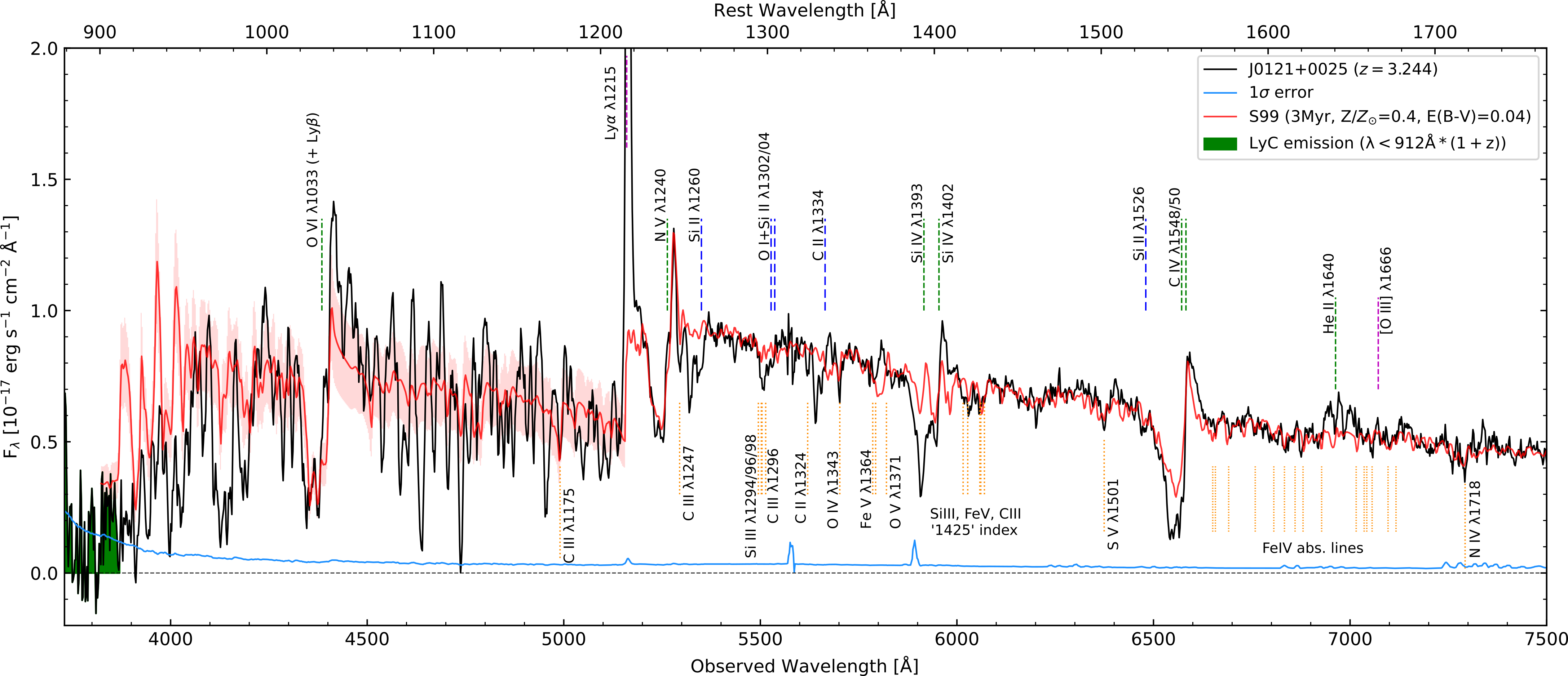}
  \caption{GTC/OSIRIS rest-frame UV spectrum of J0121$+$0025 (black) and its corresponding $1\sigma$ uncertainty (blue). Yellow dotted lines marked below the spectrum identify several photospheric absorption lines, some of them resolved and detected with high significance for which the systemic redshift was derived, $z_{\rm sys} = 3.244\pm0.001$. Low-ionization ISM absorption, wind lines in the form of P-Cygni, and nebular emission are marked in blue, green, and magenta lines, respectively. The best-fit Starburst99 model with age of 3~Myr, $Z_{\star}/Z_{\odot} =0.4$ and $E(B-V)=0.04 \pm 0.02$ is shown in red. The S99 spectrum blueward Ly$\alpha$ has been corrected for the Lyman forest absorption, using the mean and standard deviation IGM transmission $\rm T(IGM) = 0.60 \pm 0.19$ (see Section \ref{s995}). The spectral region in green corresponds to the emission below $\lambda_{0} < 912$\AA{ }related to LyC leakage.}
  \label{fig2}
\end{figure*}

Data were processed with standard {\sc Iraf}\footnote{\url{http://iraf.noao.edu/}} tasks. Each individual two-dimensional spectrum is bias-subtracted and flat-field corrected. The wavelength calibration is done using HgAr+Ne+Xe arc lamps data obtained in both days. Individual 2D spectra were background subtracted using sky regions around J0121$+$0025 ($\simeq 10^{\prime \prime}$ on both sides). The slit includes a faint ($R=23.6$, $z_{\rm phot} = 1.10 \pm0.27$) star-forming galaxy $\simeq 6.6^{\prime \prime}$ North from J0121$+$0025 (see Figure \ref{fig1}), so this region was excluded for the background statistics. The continuum of this source is detected, but not any emission line covered in the OSIRIS spectral range. Individual 1D spectra are extracted, stacked and corrected for the instrumental response using observations of the standard star Ross 640 observed in both nights. The reddening effect for the extinction in the Galaxy was taken into account adopting the extinction curve of \cite{cardelli1989} and using the extinction map of \cite{schlafly2011}. Finally, the flux of the spectrum is matched to that obtained from photometry in the $R$-band to account for slit-losses, and the final spectrum is corrected for telluric absorption using the {\sc Iraf} \textit{telluric} routine. The reduced spectrum of J0121$+$0025 is shown in Figure \ref{fig2} and presents very high signal-to-noise ratio in the continuum, $\rm SNR \simeq 20-30$ per spectral bin.

\subsection{Ancillary data}

J0121$+$0025 falls in the SDSS equatorial Stripe~82 and the rich ancillary data available in this field are used. 
These consist of optical imaging from the Hyper Suprime-Cam (HSC) on Subaru in the $G$, $R$, $Z$, and $Y$ bands from the the second data release of the HSC Subaru Strategic Program \citep{aihara2019} and MEGACAM image in the $I$ band on the Canada-France-Hawaii Telescope (CFHT), processed and stacked using the MegaPipe image staking pipeline \citep{gwyn2008}. The optical images are of great quality, both in terms of seeing conditions ($0.50^{\prime \prime}-0.85^{\prime \prime}$~FWHM) and depth, reaching $5\sigma$ limits similar or above $\simeq 25.5$ for all bands. J0121$+$0025 is detected in all images. Figure \ref{fig1} shows a cutout of J0121$+$0025 from the Subaru/HSC $R$-band. J0121$+$0025 shows a compact morphology, barely resolved ($\simeq 0.55^{\prime \prime}$ FWHM) only in the best seeing conditions image, $R$-band, with a point spread function (PSF) of $0.50^{\prime \prime}$~FWHM measured using several stars in the field. The lack of lensing structures, such as multiple images or arc-like morphologies, and the compact morphology make unlikely that J0121$+$0025 is being magnified by gravitational lensing. Using aperture photometry with a diameter of $2.5 \times \rm FWHM$, J0121$+$0025 shows a flat spectral energy distribution (SED) in the optical with magnitudes of $\simeq 21.60$ AB in the $R$- to $Y-$bands, consistent with those from SDSS photometry. Table \ref{table1} summarizes the photometry of J0121$+$0025.

\begin{table}
\begin{center}
\caption{Optical to Mid-IR Photometry of J0121$+$0025. \label{table1}}
\begin{tabular}{l c c c}
\hline \hline
\smallskip
\smallskip
Band & $\lambda_{\rm eff}$ & Magnitude & Telescope / Instrument\\
 & ($\mu$m) & (AB)  & \\
\hline 
$G$ & 0.47 & $21.98\pm 0.08$ & Subaru / HSC \\
$R$ & 0.61 & $21.60\pm 0.05$ & Subaru / HSC \\
$I$ & 0.77 & $21.57\pm 0.06$ & CFHT / MEGACAM \\
$Z$ & 0.89 & $21.53\pm 0.08$ & Subaru / HSC \\
$Y$ & 1.00 & $21.58\pm 0.08$ & Subaru / HSC \\
$J$ & 1.25 & $21.87\pm 0.23$ & VISTA / VIRCAM \\
$K_{\rm s}$ & 2.14 & $21.46\pm 0.18$ & VISTA / VIRCAM \\
$I1$ & 3.56 & $21.61\pm 0.16$ & Spitzer / IRAC \\
$I2$ & 4.51 & $22.01\pm 0.20$ & Spitzer / IRAC \\
\hline 
\end{tabular}
\end{center}
\end{table}

Near-IR imaging is also available in this field. J0121$+$0025 was observed with the VISTA InfraRed CAMera (VIRCAM) as part of the VISTA–CFHT Stripe~82 (VICS82) survey \citep{geach2017}. It is detected in the $J$- and $K_{\rm s}$-bands, with magnitudes of $21.87 \pm 0.23$ and $21.46 \pm 0.18$, respectively. 
J0121$+$0025 is also included in the Spitzer/HETDEX Exploratory Large-Area (SHELA) survey \citep{papovich2016} and is detected in the two first IRAC channels at $3.6\mu$m (I1) and $4.5\mu$m (I2) with magnitudes of $21.61 \pm 0.16$ and $22.01 \pm 0.20$, respectively. 
Finally, this region has been imaged in the X-ray by XMM-Newton with a total integration time of 2.6~ks. However, J0121$+$0025 is not detected with an X-ray flux limit of $\simeq 6 \times 10^{-15}$~erg~s$^{-1}$~cm$^{-2}$ (0.2-2 keV), corresponding to a luminosity $3\sigma$ limit of $1.2 \times 10^{45}$~erg~s$^{-1}$ at $z = 3.25$ (considering a photon index $\Gamma  = 1.7$).

\section{Results}\label{results}

\subsection{The nature of the ionizing source: SFG or AGN?}\label{sfg}

The brightness of J0121$+$0025 ($R=21.6$) and its corresponding UV luminosity at $z=3.25$, $M_{\rm UV} = -24.1$, rival that of bright QSOs at similar redshift \citep[e.g.][]{paris2018}. Therefore, it is critical to investigate first the nature of the ionizing source of J0121$+$0025.  

The high SNR rest-frame UV spectrum (Figure \ref{fig2}) is rich in absorption/emission features that are common in young starbursts, rather than in AGNs.  
Features associated with different components of the galaxy, such as stars, nebular emission and ISM, are clearly identified, and marked in Figure \ref{fig2} with different colors.

\begin{figure*}
  \centering
  \includegraphics[width=0.98\textwidth]{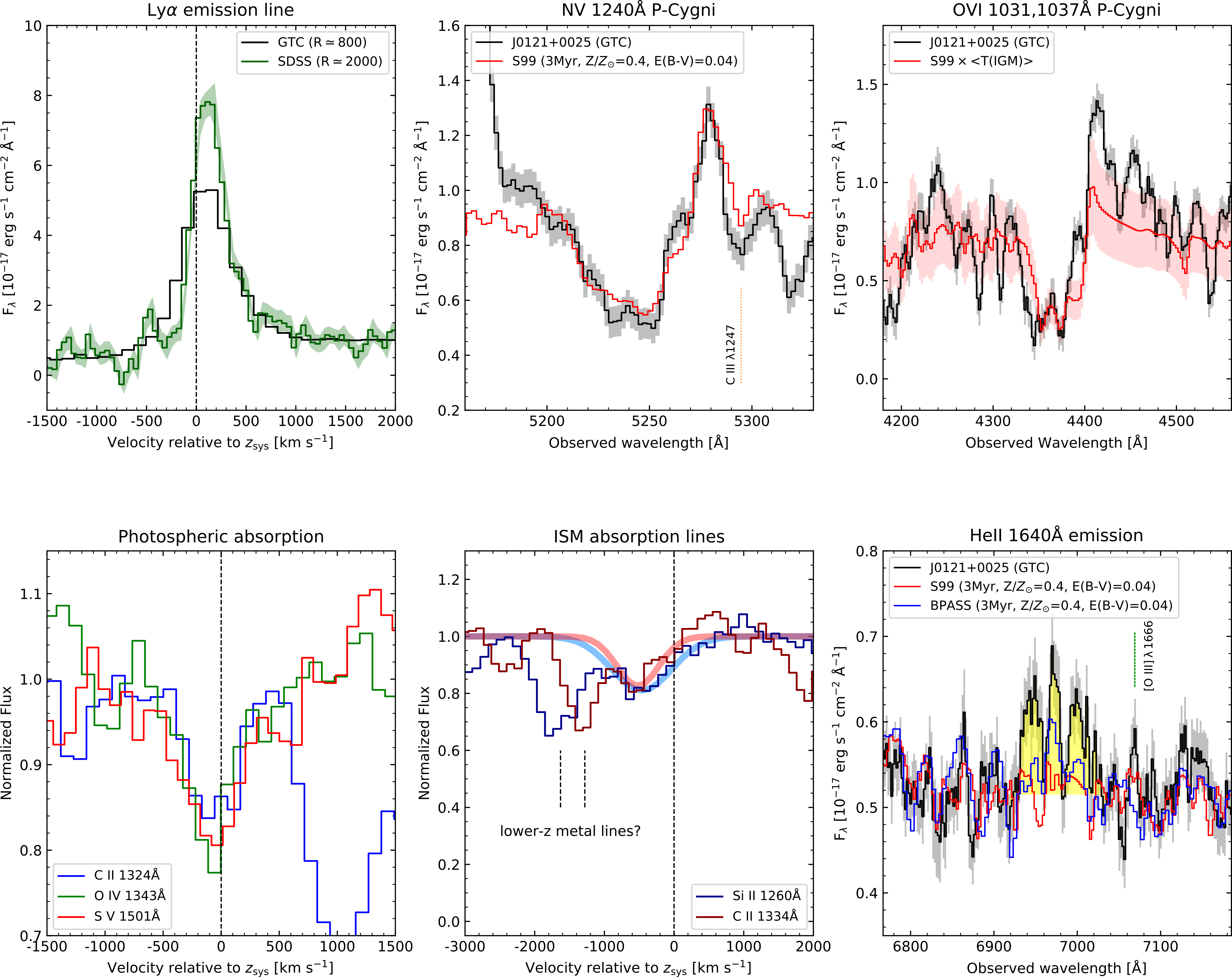}
  \caption{Spectral features detected in J0121$+$0025. Top left: Ly$\alpha$ spectral profile seen in the GTC/OSIRIS (black) and BOSS (green) spectra. It shows a narrow profile with an intrinsic $\rm FWHM = 350 \pm 40$~km~s$^{-1}$ with its peak redshifted by $\simeq 120$~km~s$^{-1}$ relative to the systemic velocity. Top middle and right panels show the N~{\sc v} and O~{\sc vi} wind lines in the form of P-Cygni (black) and the best-fit S99 model (red). Bottom left: stellar photospheric lines used to derive the systemic redshift. Bottom middle: profiles of the low-ionization ISM lines Si~{\sc ii}~1260\AA{ }and C~{\sc ii}~1334\AA{ }(blue and red, respectively). These lines are weak and have their centroids redshifted respect to the systemic velocity by $\simeq -460$~km~s$^{-1}$ and $\simeq -510$~km~s$^{-1}$, respectively. The spectrum also shows two other absorption lines (marked with dashed lines), whose nature is still unclear, but likely not physically associated with J0121$+$0025 (e.g., outflows). Bottom right: peculiar profile of the He~{\sc ii} line with three peaks (yellow). S99 and BPASS models are also shown in red and blue, respectively, but they fail to reproduce the observed emission.  }
  \label{fig3}
\end{figure*}

In particular, stellar wind P-Cygni profiles and photospheric absorption lines are detected with high significance (green and yellow dashed lines in Figure \ref{fig2}).
The detection of photospheric lines in J0121$+$0025 indicates unambiguously that the UV luminosity is dominated by stellar emission, rather than an AGN. We identify more than ten photospheric features. Some of them are resolved and detected with high significance (e.g., C~{\sc ii}~1324\AA, O~{\sc iv}~1343\AA, and S~{\sc v}~1501\AA, see Figure \ref{fig3}). We use these to determine the systemic redshift $z_{\rm sys} = 3.244\pm0.001$ of J0121$+$0025. Others are seen in blends from multiple transitions (e.g., Si~{\sc iii}~1417\AA, C~{\sc iii}~1427\AA{ }and Fe~{\sc v}~1430\AA{ }at $\lambda_{0} \simeq 1415-1435$\AA).
These stellar absorption lines are intrinsically weak in star-forming galaxies, with $EW_{0}$ typically well bellow 1\AA{ }\citep[e.g.,][]{shapley2003, steidel2016, rigby2017a}. As they are formed in the photospheres of hot stars and are seen in absorption, the background radiation should be dominated by the starlight, otherwise they would not be detected. Even a small contribution of an AGN to the UV continuum ($\lesssim 25$\%), that is featureless in these spectral regions, would make these lines disappear at the SNR of our spectrum. In addition, the observed P-Cygni profiles in N~{\sc v}~1240\AA{ }and C~{\sc iv}~1550\AA{ }can be also well explained/modelled by stellar models with a very young age ($\simeq 3$~Myr burst, see Figure \ref{fig3} and Section \ref{s99} for details), similar to those seen in other very young starbursts \citep[e.g.][]{rivera2019, vanzella2020}, some of them also very/extremely luminous \citep[][]{vanzella2018, marques2020b}. 
While some rare AGNs, such as broad or narrow absorption line QSOs (BAL/NAL QSOs), can show N~{\sc v} and C~{\sc iv} profiles mimic those of stellar P-Cygni, from the combination of a broad emission and a redshifted absorption \citep[see for example][]{bentz2004, appenzeller2005}, photospheric lines are not present in the spectra of AGNs.

The rest-frame UV morphology of J0121$+$0025 appears compact, but there is evidence of a resolved structure. Using the best seeing-condition image ($R$-band from Subaru, $0.50^{\prime \prime}$~FWHM), J0121$+$0025 appears marginally resolved with a $\rm FWHM \simeq 0.55^{\prime \prime}$, that corresponds to $\simeq 1.5-2.0$~kpc proper. Using {\sc Galfit} \citep[][]{peng2002}, the light distribution of J0121$+$0025 is better modeled using a Sersic profile instead of using a PSF model, with residuals in the region encompassing J0121$+$0025 reduced by $\simeq 50 \%$ (Figure \ref{fig1}). This suggests that the source is spatially resolved. The best-fit model (Sersic profile) gives an effective radius $r_{\rm eff} = 0.6$~pix, that corresponds to $r_{\rm eff} = 0.1^{\prime \prime}$ assuming the Subaru pixel scale  $0.168^{\prime \prime}$/pix, and $r_{\rm eff} \sim 0.8$~kpc at $z=3.244$. 
Although it has been shown that {\sc Galfit} can recover effective radius down to $\sim 0.5$~pix if the PSF is properly known and the source is bright enough (see: \citealt{vanzella2017}), we assume conservatively an $r_{\rm eff} < 1$~kpc.

From another perspective, the Ly$\alpha$ line shows a narrow profile with an intrinsic FWHM $\simeq 350$~km~s$^{-1}$ compatible with star-forming galaxies (see Figure \ref{fig3} and Section \ref{s993}). The AGN population typically shows broad Ly$\alpha$ profiles, up to several hundreds or thousands km~s$^{-1}$, even in the case of narrow line AGNs \citep[e.g.,][]{hainline2011b}. In addition, the observed Ly$\alpha$ equivalent width, flux and corresponding luminosity can be well explained by star-formation only (discussed in Section \ref{s99}), without the need to invoke an AGN contribution. 
He~{\sc ii}~1640\AA{ }is also detected, showing a broad profile with $\simeq 2500$~km~s$^{-1}$. Because He~{\sc ii} is a non-resonant (and recombination) line, its origin is likely stellar and not nebular (from an AGN), otherwise we would expect to detect a similar or even broader profile in the resonant Ly$\alpha$ line, which is not the case. This will be discussed in more detail in Section \ref{s99}. Nebular emission of [O~{\sc iii}]~1666\AA{ }is also detected (with low significance) and shows a narrow profile, not resolved in our OSIRIS spectrum ($<450$~km~s$^{-1}$, Figures \ref{fig2}). This line is not present in the spectra of typical AGNs.

\begin{figure}
  \centering
  \includegraphics[width=0.47\textwidth]{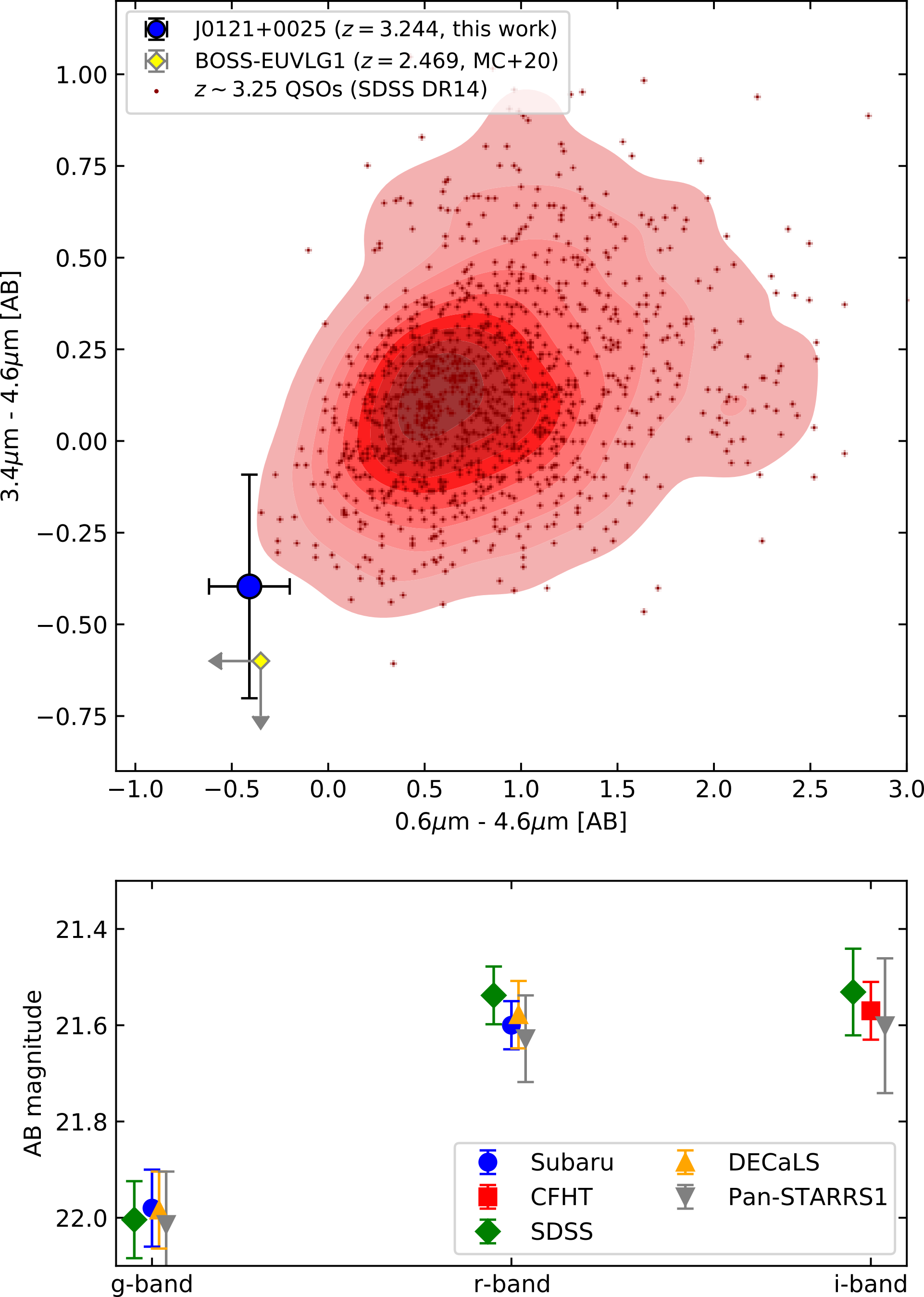}
  \caption{Top: Comparison between the 3.4$\mu$m - 4.6$\mu$m and 0.6$\mu$m - 4.6$\mu$m colors of J0121$+$0025 (blue) and those from SDSS/BOSS QSOs \citep{paris2018} at similar redshift (red). J0121$+$0025 shows bluer colors than typical QSOs. For comparison, the young starburst BOSS-EUVLG1 ($\simeq$4~Myr, \citealt{marques2020b}) is also shown in yellow. Bottom: observed magnitudes of J0121$+$0025 in various filter bands ($g$, $r$ and $i$) with different telescopes and epochs. No variability is detected in J0121$+$0025.}
  \label{fig2p2}
\end{figure}

The presence of a UV-faint or an obscured type-II AGN is more difficult to exclude. Unfortunately, we cannot use rest-frame UV and optical line diagnostics \citep[e.g.,][]{BPT, nakajima2018} to discriminate between star-formation and AGN, as these lines are not covered by the OSIRIS spectrum. 
However, the mid-IR photometry of J0121$+$0025 disfavours the presence of such AGN contribution (top panel of Figure \ref{fig2p2}). J0121$+$0025 shows a flat/blue spectral energy distribution (SED) from the rest-frame UV to near-IR with $R-I2 = -0.41\pm0.21$ and $I1 - I2 = -0.40\pm0.31$, where $R$, $I1$ and $I2$ bands probe rest-frame wavelengths of $\simeq 0.16\mu$m, $0.84\mu$m and $1.06\mu$m, respectively. As shown in Figure \ref{fig2p2}, the optical to mid-IR colors of J0121$+$0025 place it far away from the locus of AGNs at similar redshift \citep{paris2018}. AGNs tend to have red optical-to-mid-IR SEDs due to the rising emission at $\lambda_{0} \gtrsim 1\mu$m (rest) originated by the dust torus \citep[][]{assef2013}. 
We also check for possible variability from an AGN in the optical photometry. In the bottom panel of Figure \ref{fig2p2} we compare the observed magnitudes in the $g$, $r$ and $i$ bands of J0121$+$0025 from Subaru, CFHT, SDSS, DECaLS and Pan-STARRS1, that probe different epochs, from MJD 51544 to 58362 ($\sim 18$~years). No variability is detected in J0121$+$0025.
Lastly, J0121$+$0025 is not detected in X-rays in the 2.6~ks XMM-Newton data, although the corresponding $3\sigma$ limit $L_{\rm (0.2-2)keV} = 1.2 \times 10^{45}$~erg~s$^{-1}$ is not deep enough to further explore a possible X-rays emission of an AGN. Note that a significant X-ray emission from star-formation is still expected in J0121$+$0025. Assuming $\rm SFR = 981 M_{\odot}$~yr$^{-1}$ (see Section \ref{s996}) and following \cite{grimm2003}, we expect an X-ray emission from star-formation $L_{\rm x} (\rm SFR) \sim 10^{43}$~erg~s$^{-1}$.

Overall, the shape of the rest-frame UV spectrum, the detection of stellar features (photospheric absorption and wind lines), the resolved morphology and the multi-wavelength SED highly suggest that the luminosity of J0121$+$0025 is being powered by a vigorous starburst and that a significant contribution of an AGN at these wavelengths is unlikely.

\subsection{Rest-frame UV properties} \label{s99}

\subsubsection{Young stellar population: age, metallicity and attenuation}\label{s991}

One of the most prominent features in the spectrum of J0121$+$0025 is the P-Cygni associated with wind lines, that are much stronger than in typical LBGs \citep[e.g.,][]{shapley2003}. These lines are produced by strong outflows of material from the most massive stars, whose strength and spectral shapes depend strongly on the age and metallicity (and on the initial mass function, IMF) of the stellar population, being N~{\sc v} and C~{\sc iv} by far the most sensitive features \citep[][]{chisholm2019}.

To infer the properties of the young stellar population in J0121$+$0025, the observed line profiles of N~{\sc v} and C~{\sc iv} are compared to those obtained with the spectral synthesis code {\sc Starburst99} \citep[S99:][]{leitherer1999}, following the same methodology described in \cite{marques2018} and \cite{marques2020}. We use S99 instead of BPASS models \citep{stanway2016}, because the later are less able to match the details of the P-Cygni absorption of C~{\sc iv} (see discussion and Figure 5 in \citealt{steidel2016}). Briefly, we generate high-resolution (0.4\AA) UV spectra using standard Geneva tracks with a grid of metallicities ($Z_{\star}/Z_{\odot}$, where $Z_{\odot}=0.02$) of 0.05, 0.2, 0.4 and 1, and burst ages from 1~Myr to 30~Myr. An IMF with a power slope index $\alpha=-2.35$ over the mass range $0.5 < M_{\star}/M_{\odot} < 100$ is considered. S99 outputs are redshifted to $z=3.244$, smoothed to the spectral resolution of the OSIRIS spectrum and rebined to the spectral bin of 2.12\AA. Dust attenuation is also taken into account, considering values $E (B-V)_{\star}$ ranging from 0 to 0.2 and using the \cite{calzetti2000} extinction curve and its extension to short wavelengths ($<0.15\mu$m) provided by \cite{reddy2016a}. Spectral windows that are \textit{free} of of absorption/emission features \citep{rix2004} and strong sky-subtracted residuals are used to offset the flux of S99 models.
We then compare the observed N~{\sc v} and C~{\sc iv} profiles with those from S99, performing a $\chi^{2}$ minimization over the spectral range $1225-1245$\AA{ }for N~{\sc v} and $1528-1538$ for C~{\sc iv}, excluding in the fit spectral regions that could be affected by interstellar absorption or nebular emission, which is particularly relevant for C~{\sc iv}. 

Wind lines of N~{\sc v} and C~{\sc iv} are well reproduced by a $Z_{\star}/Z_{\odot} \simeq 0.4$ and $ \simeq 3$~Myr burst of star formation (red in Figure \ref{fig2}). A color excess of the stellar continuum $E (B-V)_{\star}=0.04\pm0.02$ is also inferred, which is compatible with the observed UV slope, $\beta_{\rm UV} = -2.05 \pm 0.10$. Scaling the best-fit S99 model to $M_{\rm UV}=-24.1$, this leads to a burst mass log($M_{\star}/M_{\odot})=9.8$. According to S99, the number of O-type stars is $\sim 8 \times 10^{5}$ yielding an intrinsic ionizing photon production rate $N_{\rm int} \rm (LyC) \simeq 1.4 \times 10^{55}$~s$^{-1}$ and the production efficiency, $\xi = N_{\rm int}$~(LyC) / $L_{\rm UV, int}$, is log($\xi ) = 25.2$. 

Considering the full spectral range, the overall agreement between the best-fit model and the observed spectrum is mixed. Some stellar features are well-fitted, as the N~{\sc v} and C~{\sc iv} P-Cygni and some photospheric absorption, while others show poor agreement. The profile of Si~{\sc iv}~1393,1402\AA{ }shows evidence of a P-Cygni contribution (in addition to the ISM absorption), but the model underpredicts it, which could suggest an additional contribution of a slightly older stellar population with age $\simeq 5$~Myr \citep[see Figure 6 in][]{leitherer2001}. It is also interesting to note that the OSIRIS spectrum shows evidence of a P-Cygni around the photospheric blanketing by Fe~{\sc v} and O~{\sc v} around $\simeq 1360-1375$\AA, but it appears slightly blueshifted with respect to the predicted S99 model, by $\simeq 400-500$~km~s$^{-1}$. On the other hand, other stellar features are relatively well reproduced, such as the region around $\simeq 1420-1430$\AA{ }from the blended emission of Si~{\sc iii}, Fe~{\sc v} and C~{\sc iii} transitions (also called the "1425" index), the photospheric S~{\sc v}~1501\AA{ }line, and the Fe~{\sc iv} complex around $\sim 1600$\AA{ }(see Figure \ref{fig2}). 
The model also predicts a relatively strong P-Cygni in O~{\sc vi}~1031,1033\AA, but is still weak compared to the observed one (Figure \ref{fig3}), which could indicate an even younger stellar population ($\leq 2$~Mry). Note however that the profile of O~{\sc vi} is likely affected by the contribution of the Ly$\beta$ absorption and the IGM attenuation, that could impact the observed profile.

Overall, the best-fit S99 model with $Z_{\star}/Z_{\odot} = 0.4$, age of 3~Myr and $E (B-V)_{\star}=0.04$ fits reasonably well the observed spectrum of J0121$+$0025, in particular the wind lines N~{\sc v} and C~{\sc iv}, that are the most sensitive features to these parameters \citep[][]{chisholm2019}. Note however that the inferred $Z_{\star}$ and age are model-dependent and should be treated with caution and considered as approximated values. This is particularly relevant for the metallicity, which is less constrained, as our analysis is limited to discrete models with a set of four metallicities. Consequently we can only say that the model with $Z/Z_{\odot} = 0.4$ is favoured with respected to the other models ($Z/Z_{\odot} = 0.05,0.2,1$). 
In addition, we are assuming a single burst model, which might not be realistic (nor a continuum star-formation rate). Nevertheless, the age is much better constrained and should be $<8$~Myr, otherwise the wind line of N~{\sc v} would appear much weaker ($\simeq 10$~Myr) or almost non-existent ($> 10$~Myr). The same applies if a continuous star-formation history is assumed. In this case, the redshifted emission of N~{\sc v} could be well described with a continuous SFH with an age up to $\simeq 20$~Myr, but not the corresponding blueshifted absorption, that would be underestimated for ages $\gtrsim 10$~Myr due to the increasing contribution of B-type stars to the UV continuum. In addition, the blue mid-IR color $I1 - I2 = -0.4 \pm 0.31$ highly supports a very young age of the stellar population, that would be difficult to explain with a sightly older stellar population ($\gtrsim 15-20$~Myr) with a continuous star-formation rate.

\subsubsection{Broad He~{\sc ii} emission}\label{s992}

The He~{\sc ii}~1640\AA{ }emission is shown in detail in the bottom right panel of Figure \ref{fig3} and presents a complex morphology characterized by a broad profile with two absorption in the central part of the line forming a triple peaked emission. Fitting a Gaussian profile and excluding the two absorption features in the fit, we measure a $\rm FWHM \simeq 2500$~km~s$^{-1}$. Because of its broadness, this emission is likely stellar in origin, otherwise we would expect a similar or even broader profile in other nebular lines like the resonant Ly$\alpha$ line. However, this is not the case as Ly$\alpha$ shows a narrow profile ($\simeq 350 $~km~s$^{-1}$~FWHM, see Section \ref{s993}). The nebular [O~{\sc iii}]~1666\AA{ }line is also detected and show an unresolved profile in the OSIRIS spectrum ($<450$~km~s$^{-1}$~FWHM, see Figure \ref{fig2}). 

We measure a rest-frame equivalent width $EW_{0}=3.2 \pm 0.3$\AA{ }for He~{\sc ii} (corresponding to the yellow region of Figure \ref{fig3}), that is much larger than that inferred in the $z\sim 3$ LBG composite spectrum of \cite{shapley2003} ($EW_{0}^{\rm LBGs} = 1.3 \pm 0.3$). 
The He~{\sc ii} emission in J0121$+$0025 is also stronger than the average $EW_{0} \simeq 2.5$\AA{ }found in some extreme Wolf-Rayet (WR) star clusters in the local Universe (\citealt{chandar2004}, although see the cases of NGC~3125-1 with $EW_{0} \simeq 7.4$\AA, or the dwarf galaxy II~Zw~40 with $EW_{0} \simeq 7.1$\AA, \citealt{leitherer2018}).
As shown in Figure \ref{fig3}, the best fit S99 model clearly underpredicts the strength of the observed He~{\sc ii} profile. In fact, \cite{brinchmann2008} predict a $EW_{0} \rm (S99) =0.3$\AA{ }for a S99 burst model with $Z_{\star}/Z_{\odot}=0.4$ and $3$~Myr age. Even considering an extreme case with a continuous star-formation history and $Z_{\star} = Z_{\odot}$, S99 models predict $EW_{0} \rm (S99) \leq 2.4$\AA{ }\citep[see Figure 2 of][]{brinchmann2008}. For comparison, a BPASS binary model \citep{stanway2016} with the same age and metallicity is also shown in Figure \ref{fig3}, and although it predicts stronger stellar He~{\sc ii} emission than S99 models (see also: \citealt{steidel2016}), it still under-predicts the observed emission in J0121$+$0025.

The strength of He~{\sc ii} in J0121$+$0025 raises now the question regarding the presence of more exotic stellar populations. For example, He~{\sc ii} appears very strong in the spectra of very massive stars (VMS, $>100 M_{\odot}$) in the central cluster R136 of the 30 Doradus star-forming region \citep[see:][]{crowther2016}. In fact, the contribution of such massive stars has been proposed to explain the large $EW_{0}$'s in He~{\sc ii} in a few local star-forming galaxies (with $EW_{0} \simeq 2.0 - 4.7$\AA, \citealt{senchyna2021}). Interestingly, the complex triple-peaked He~{\sc ii} profile seen in the spectrum of J0121$+$0025 resembles that observed in two star-forming galaxies analysed by \cite{senchyna2021}, namely SB~179 and SB~191 (their Figure 6), that according to these authors could be the product of rapid rotation of rare Onfp stars \citep[e.g.,][]{walborn2010}. 
Investigating in detail the nature of the He~{\sc ii} line and  WR and VMS content (and possible nebular contribution) is out of the scope of this work as it requires follow-up observations (e.g., high-spectral resolution of the He~{\sc ii} line and near-IR spectroscopy to put strong constraints on the metallicity), and in addition updated stellar models with the inclusion of very massive stars.

\subsubsection{The Ly$\alpha$ line}\label{s993}

The Ly$\alpha$ line in J0121$+$0025 shows a spectrally unresolved profile in the OSIRIS spectrum ($R\simeq 800$), but it is slightly resolved in the higher resolution BOSS spectrum (Figure \ref{fig3}). Fitting a Gaussian profile, we measure an intrinsic $\rm FWHM = 350 \pm 40$~km~s$^{-1}$, after correcting for the instrumental broadening ($\simeq 150$~km~s$^{-1}$), and a rest-frame equivalent width $EW_{0} \rm (Ly\alpha) = 14 \pm 3$\AA. The Ly$\alpha$ line has its peak closed to the systemic redshift, redshifted by $v_{\rm peak} \simeq 120\pm50$~km~s$^{-1}$. 
It is still not clear why the measured $EW_{0} \rm (Ly\alpha)$ appears so low, compared to the intrinsic $EW_{0}^{\rm int} \rm (Ly\alpha) \sim 100$\AA{ }expected for a $\simeq 3$~Myr age burst and $Z_{\star}/Z_{\odot}$ \citep{schaerer2003}. Roughly, this yields a Ly$\alpha$ escape fraction of $\sim 14\%$, which is lower than $f_{\rm esc, abs} \rm (LyC) \approx 40\%$ (see Section \ref{s995}). Considerable fiber/slit losses, strong IGM attenuation near $\lambda_{0} \simeq 1215$\AA, destruction of Ly$\alpha$ photons by dust, and/or large $f_{\rm esc}$ of ionizing photons could in principle explain such differences. In addition, there could be more ionising photons available than H~{\sc i} to be ionised, i.e., the ISM in J0121$+$0025 could be mostly ionized approaching to a density-bounded geometry. Such a scenario should be further investigated using, e.g., the [O~{\sc ii}]~3727,29\AA{ }and [O~{\sc iii}]~5008\AA{ }lines ([O~{\sc iii}]/[O~{\sc ii}] ratio). In addition, the H$\beta$ line could provide constraints on the properties of the Ly$\alpha$ emission, both in terms of its intensity and spectral shape. These lines ([O~{\sc ii}], [O~{\sc iii}] and H$\beta$) are redshifted to the near-IR $H$- and $K$-bands and are accessible from the ground.

Using the BOSS/SDSS spectrum, that is less affected by flux losses, we measure a total Ly$\alpha$ flux of $F \rm (Ly\alpha) = (5.72 \pm 0.10) \times 10^{-16}$~erg~s$^{-1}$~cm$^{-2}$. This corresponds to a luminosity log($L_{\rm Ly\alpha}/$erg~s$^{-1})= 43.8\pm 0.1$ at the redshift of J0121$+$0025. 
To test whether or not this luminosity could be explained by star-formation, we compare the SFR of J0121$+$0025 obtained from SED fitting ($\rm SFR (SED) =981 \pm 232 M_{\odot}$~yr$^{-1}$, see Section \ref{s996}) to the Ly$\alpha$ star-formation rate using the \cite{kennicutt1998} conversion. Assuming case-B recombination and the \cite{salpeter1955} initial mass function (IMF), the Ly$\alpha$ luminosity corresponds to a $\rm SFR (\rm Ly\alpha) \simeq 80$~$M_{\odot}$~yr$^{-1}$. Even considering $f_{\rm esc} \rm (LyC) >0$ (and so $\rm SFR (\rm Ly\alpha) \gtrsim 80$~$M_{\odot}$~yr$^{-1}$), star-formation can naturally explain the observed Ly$\alpha$ flux and the corresponding luminosity.

\subsubsection{ISM, kinematics and covering fraction}\label{s994}

In addition to the strong Ly$\alpha$ line and other stellar features, the high S/N spectrum reveals also absorption features that are associated with the interstellar medium gas (ISM) and produced by the resonance transition of several ionic species. 

Low-ionization ISM lines (LIS) of Si~{\sc ii}~1260\AA, C~{\sc ii}~1334\AA{ }and Si~{\sc ii}~1526\AA{ }are detected in the spectrum. Absorption in O~{\sc i}~1302\AA{ }and Si~{\sc ii}~1304\AA{ }is also detected but it is not resolved and, in addition, its profile is also contaminated by the photospheric complex around $\lambda_{0} \simeq 1294-1298$\AA. 
Others LIS lines that are usually present in the spectra of LBGs, such as Fe~{\sc ii}~1608\AA{ }or Al~{\sc ii}~1670\AA{ } \citep{shapley2003}, are not detected in J0121$+$0025.

Since LIS lines are seen against the continuum provided by starlight of J0121$+$0025, they are useful to probe the kinematics of the gas along the line-of-sight. Despite the low spectral resolution of the OSIRIS spectrum, the centroids of the LIS lines are clearly blueshifted with respect to the systemic redshift, by $v_{\rm peak} \rm (LIS) \simeq -450$~km~s$^{-1}$ (Figure \ref{fig3}). 
This could be an indication of fast gas outflows caused by supernova explosions and stellar winds, similar to those detected in the other extremely UV luminous starburst BOSS-EUVLG1 ($v_{\rm peak} \rm (LIS) \simeq -400$~km~s$^{-1}$ \citealt{marques2020b, alvarez2021}) or in the Sunburst LyC emitter \citep{rivera2017, vanzella2021}.

\begin{figure*}
  \centering
  \includegraphics[width=0.98\textwidth]{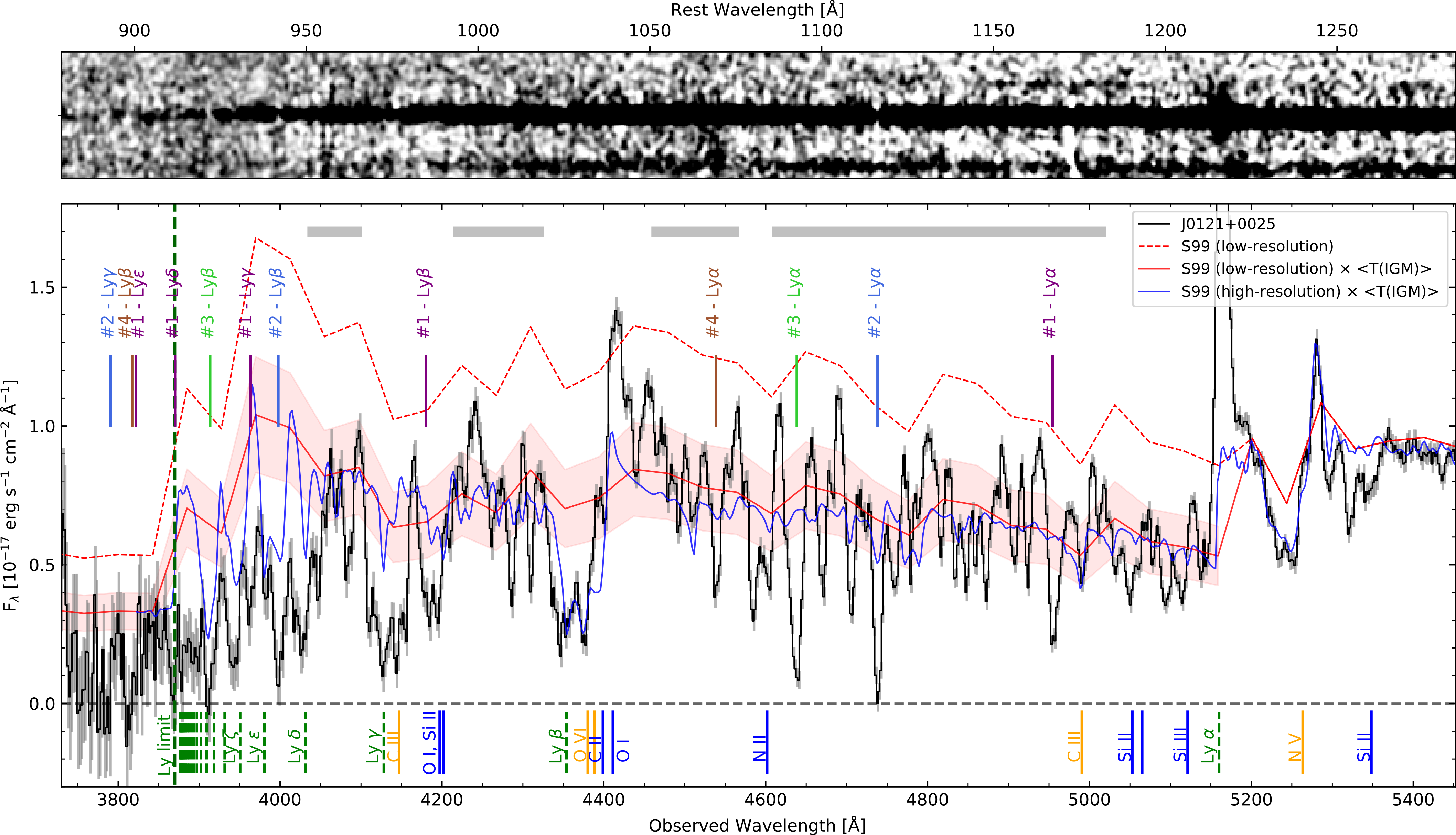}
  \caption{2D (top) and 1D (bottom) spectra of the far-UV region of J0121$+$0025. The 2D spectrum has been smoothed for visualization purpose. The GTC spectrum is in black and its corresponding $1\sigma$ uncertainty is in grey. Features associated with the Lyman series (from Ly$\alpha$ to the Lyman limit), ISM and stellar absorption (photospheric and wind lines) are marked below the spectrum in green, blue, and yellow colors, respectively. The low-resolution best-fit S99 model (3~Mry age, $Z_{\star}/Z_{\odot}=0.4$ and $E(B-V)_{\star} = 0.04$) is plotted in red, one corrected for the IGM transmission (<$T(IGM)$> $=0.60 \pm 0.19$, solid red) and other assuming $T(IGM)=1$ (dashed red). The high-resolution S99 model is also shown in blue. Horizontal grey lines mark the spectral windows used to infer $T(IGM)$ (see text). These exclude the regions associated with the Lyman series and ISM from J0121$+$0025, so that we probe H~{\sc i} absorbers from the Lyman forest in the line of sight only. Vertical lines above the spectrum mark the position of four strong H~{\sc i} absorbing systems identified at $z =  3.076, 2.898, 2.816$ and 2.733 (\#1 to \# 4, respectively). We also note that the 2D spectrum also shows a faint continuum offset by $\simeq 6^{\prime \prime}$ of J0121$+$0025 from a low-$z$ star-forming galaxy (with a $z_{\rm phot}=1.10$, see Figure \ref{fig1}).} 
  \label{fig5}
\end{figure*}

The OSIRIS spectrum also shows two additional absorption lines at $\simeq 5320$\AA{ } and $\simeq 5640$\AA{ }(Figure \ref{fig3}), whose origin is still not clear. These lines appear close to Si~{\sc ii}~1260\AA{ }and C~{\sc ii}~1334\AA{ }of J0121$+$0025, by $\simeq -1700$~km~s$^{-1}$ and $\simeq -1200$~km~s$^{-1}$, respectively (Figure \ref{fig3}). These lines could be associated with outflows from J0121$+$0025, but such scenario is unlikely. Because these lines have broadly similar ionization potentials, the outflowing gas traced by Si~{\sc ii} and C~{\sc ii} should be kinematically coherent (and co-spatial), which is not the case. Moreover, the profile of Si~{\sc ii}~1526\AA{ }does not show this secondary absorption component. It is possible that the absorption line at $\simeq 5320$\AA{ }is associated with Si~{\sc iv}~1393\AA{ }at $z=2.816$, for which Ly$\alpha$ is also detected in absorption around $\simeq 4640$\AA{ }(H~{\sc i} absorbing system \#3 in Figure \ref{fig5}). In such a case, the corresponding Si~{\sc iv}~1402\AA{ }absorption of this intervening system is contaminating the profile of Si~{\sc ii}~1260\AA{ }of J0121$+$0025. For the other absorption at $\simeq 5640$\AA, we are not able to find any redshift solution, so the redshift of this intervening system is still not known.

Regarding the strength ($EW_{0}$) of LIS lines in J0121$+$0025, we fit Gaussian profiles to the absorption profiles, excluding the contribution of the secondary, not related absorption component mentioned before (Figure \ref{fig3}). We measure rest-frame equivalent widths of $1.05\pm0.15$\AA, $0.59\pm0.08$\AA, and $0.81\pm0.10$\AA, for Si~{\sc ii}~1260\AA, C~{\sc ii}~1334\AA{ }and Si~{\sc ii}~1526\AA, respectively. Note that the measured equivalent width of Si~{\sc ii}~1260\AA{ }should be considered as an upper limit, because its profile is likely contaminated by the intervening metal lines Si~{\sc iv}~1402\AA{ }at $z=2.816$. These lines appear spectrally resolved with an intrinsic $\rm FWHM \simeq 550-650$~km~s$^{-1}$. 
For comparison, the $z\sim 3$ LBG composite spectrum of \cite{shapley2003} shows large $EW_{0}$'s for the same lines, with $EW_{0} \simeq 1.7$\AA. 

The weakness of LIS lines in J0121$+$0025 could arise either by a low geometric covering fraction of the gas, $C_{f}$, a low ion column density, and/or a highly ionized ISM. Considering the linear part of the curve of growth (i.e. low column density), the ratios of the $EW_{0}(1260)/EW_{0}(1526)$ can be related through their oscillator strengths, for which we would expect a $EW_{0}(1260)/EW_{0}(1526) \simeq 5$ if these lines were not saturated. However, we measure $EW_{0}(1260)/EW_{0}(1526) \lesssim 1.3$, suggesting that at least one of these lines is saturated. This is not surprising at all, as these lines appear almost always saturated in the spectra of star-forming galaxies, even in damped-Ly$\alpha$ systems with sub-solar metallicities \citep[e.g.,][]{des2006}. 

Taking this in consideration, it is possible that the weakness of LIS lines arises from a low $C_{f}$. Assuming the optically thick regime and an ionization-bounded ISM with a uniform dust-screen  geometry, the $C_{f}$ can be inferred using the residual intensity of the absorption line, $I$, so that $C_{f} = 1 - I/I_{0}$, where $I_{0}$ is the continuum level. We measure $I/I_{0} \simeq 0.8$ for Si~{\sc ii}~1260\AA{ }and C~{\sc ii}~1334\AA, yielding to $C_{f} \rm (SiII) \simeq 0.2$. Note that with the low spectral resolution of our data we are likely overestimating $I/I_{0}$, however this effect should not be dominant as the lines are spectrally resolved.
Following \cite{gazagnes2018}, this yields a neutral gas covering fraction $C_{f} \rm (HI) \simeq 0.55$, for that a significant fraction of LyC photons could escape. In fact, using the prescriptions of \cite{chisholm2018} (see also Saldana-Lopez in prep.), the inferred $C_{f} \rm (HI) \simeq 0.55$ leads to a predicted $f_{\rm esc, abs}^{\rm pred} \rm (LyC) \approx 0.25$, which is consistent with the observed value from the spectrum (next section).

\subsection{Lyman Continuum radiation}\label{s995}

\subsubsection{The direct detection of LyC}

One of the most remarkable features observed in the OSIRIS spectrum of J0121$+$0025 is the detection of emission below $\lambda_{\rm obs} \simeq 3880$\AA, i.e. $\lambda_{\rm 0} < 911.8$\AA, that can be related to LyC leakage. 
This emission is real and it is not related with detector artifacts (e.g., cosmic rays, flat-field correction) or bad sky subtraction. It is detected in the spectra of different nights, as well as in the much shallower BOSS spectrum although with less SNR ($\simeq 2.3 \sigma$).

A zoom-in into this region is shown in Figure \ref{fig5}. The observed flux emission at rest-frame $880 - 910$\AA{ }has a total SNR of $7.9$ and an average SNR per spectral bin of $0.98$. 
The mean flux density in this spectral region is $f_{900} (\rm obs) = 0.781\pm0.099 \mu$Jy, which corresponds to a magnitude of $m_{900} = 24.17$~(AB). For comparison, the mean non-ionizing UV flux density, estimated from the OSIRIS spectrum over the rest-frame range $1490 - 1510$\AA, is $f_{1500} = 8.874\pm0.399 \mu$Jy. Combining these measurements, we find a ratio of the ionizing to non-ionizing flux density $(f_{900} / f_{1500})_{\rm obs} = 0.088 \pm 0.012$, which corresponds to a $\Delta m = 2.64$~(AB).
It is worth noting that the ionizing emission $\lambda_{\rm obs} < 3880$\AA{ }suffers for an apparent absorption at $\lambda_{\rm obs} \simeq 3813$\AA. After careful inspection, this absorption could be related with Ly$\beta$ and Ly$\delta$ absorption associated with two H~{\sc i} systems at $z = 2.733$ and $z=3.076$, respectively (\#1 and \#4 in Figure \ref{fig5}). 

To infer the relative LyC photon escape fraction, we compare the observed ionizing flux density of J0121$+$0025 at $900$\AA, $f_{900} (\rm obs)$, to that from the S99 model, $f_{900} (\rm S99)$, that best represents the shape of the rest-frame UV spectrum (Section \ref{s991}), i.e., a burst with an age of 3~Myr, $Z_{\star}/Z_{\odot} = 0.4$ and $E(B-V)=0.04$ (Figures \ref{fig3} and \ref{fig5}). The S99 model has been already corrected for $E(B-V)$, thus it incorporates the relative attenuation between 900\AA{ }and 1500\AA{ }due to the variation of the attenuation coefficient with the wavelength ($k_{900} \simeq 14.3$ and $k_{1500} \simeq 10.3$, assuming the \citealt{calzetti2000} curve and its extension to the UV provided by \citealt{reddy2016a}).\footnote{The inferred $E(B-V)$ using \cite{calzetti2000} curve is already low, so using other extinction curves (e.g., SMC) will have little impact in our results.} To probe the region below $\lambda_{0} < 912$\AA, we use the low-resolution version of S99 model because it extends to the LyC region. Finally, we also take into account the contribution of the IGM transmission, $T(IGM)$. The relative LyC photon escape fraction, $f_{\rm esc, rel} (LyC)$, can thus be expressed using the following formulation:

\begin{equation}\label{eq1}
    f_{\rm esc, rel} (LyC) = \frac{f_{900} (obs)} {f_{900} (S99)} \times \frac{1}{T(IGM)}.
\end{equation}

Both $f_{900} (\rm obs)$ and $f_{900} (\rm S99)$ are measured using the same spectral window, defined at $880 - 910$\AA{ }in the rest-frame. We find $f_{900} (\rm obs) / f_{900} (\rm S99) = 0.34 \pm 0.04$, where the uncertainties arise from $f_{900} (\rm obs)$. We note that Equation \ref{eq1} is consistent to that used in the literature \citep[e.g.,][]{shapley2016, vanzella2016, steidel2018}, where $f_{900} / f_{1500}$ is used instead of $f_{900}$, because in our case $f_{1500} \rm (S99)$ has been already matched to $f_{1500} \rm (obs)$ (see Figure \ref{fig2}), thus  $f_{1500} \rm (obs) \simeq$~$f_{1500} \rm (S99)$. 

A precise estimate of $f_{\rm esc, rel} (LyC)$ is not possible given the stochastic nature of $T(IGM)$ and the large fluctuation of the attenuation in one single line-of-sight \citep{inoue2008, inoue2014}. However, it is still possible to place rough constraints. Assuming that $f_{900} (\rm obs) / f_{900} (\rm S99) = 0.34 \pm 0.04$ is well constrained,\footnote{For an age of 3~Myr the predicted ionizing flux at 900\AA{ }using S99 and BPASS (including binaries) are roughly the same, see e.g., \cite{chisholm2019}. } this implies that $T(IGM)$ should be at least larger than $>0.34$ to keep a physical $f_{\rm esc, rel} (LyC) < 1$ \citep[e.g.,][]{vanzella2012}. On the other hand, $f_{\rm esc, rel} (LyC)$ must be $\geq 0.34$, where the most extreme value ($0.34$) stands for a completely transparent IGM.

To get a more quantitative estimate of $T(IGM)$, we use the non-ionizing part of the OSIRIS spectrum, from $912 - 1215$\AA, and compare it to that of S99 best-model. We exclude spectral regions that are associated with the Lyman series and ISM from J0121$+$0025 (marked in Figure \ref{fig5}) that are not included in the S99 models, so that we probe H~{\sc i} absorbers from the Lyman forest in the line of sight only, and not the the CGM and ISM associated to J0121$+$0025. The spectral regions used to estimate $T(IGM)$ are marked as horizontal grey lines in Figure \ref{fig5}. Note however that with this comparison, we are inferring the $T(IGM)$ in this spectral range, not necessarily at $\leq 912$\AA, still it can serve as a first-order approximation. It is worth noting that the S99 model predicts a relatively strong break around the Lyman limit, which is not observed in the spectrum of  J0121$+$0025 (and in other LyC leakers, see e.g., \citealt{steidel2018}).
We find a mean value and standard deviation $T(IGM) = 0.60 \pm 0.19$, which is compatible with the \cite{inoue2014} model or that obtained in other LyC leakers at similar redshifts and $f_{\rm esc, rel} (LyC)$ \citep[e.g.,][]{shapley2016} using Monte Carlo simulations of the IGM transmission, but it is larger than those obtained by, e.g., \cite{steidel2018} and \cite{fletcher2019}, with mean values of $T(IGM) \simeq 0.3$.

Using Equation \ref{eq1} we infer $f_{\rm esc, rel} (LyC) \sim 0.56$, with possible values ranging from $0.34$ to $1$. Note that we are considering only the uncertainties due to the IGM. A sightly older stellar population, for example a burst with $\simeq 6$~Myr, will produce less ionizing photons than the model we are considering ($3$~Myr) by a factor of $\simeq 2$. However, the $6$~Myr burst model would require less extinction to explain the observed $\beta_{\rm UV} = -2.05$, so that the $f_{900} (\rm obs) / f_{900} (\rm S99)$ would be roughly similar in both cases. Other sources of uncertainty can impact the inferred $f_{\rm esc, rel} (LyC)$, but are extremely difficult to quantify. These include the uncertainties due to the flux calibration at the blue edge of the OSIRIS spectrum, which has a total efficiency $\lesssim 5\%$ at $\lambda < 4000$\AA{  }only,\footnote{\url{http://www.gtc.iac.es/instruments/osiris/\#Spectroscopic_Photon_Detection_Efficiency}} or differential slit-losses between $\lambda_{0} < 912$\AA{ }and $\lambda_{0} \simeq 1500$\AA, due to the effect of the atmospheric dispersion and the broadening of the point-spread function at blue wavelengths. Nevertheless, the uncertainties due to the IGM should be dominant.
Given this, the absolute LyC escape fraction, $f_{\rm esc, abs} (LyC)$, defined as \cite[e.g.,][]{leitet2013}:

\begin{equation}\label{fesc_abs}
    f_{\rm esc, abs} (LyC) =  f_{\rm esc, rel} (LyC) \times 10^{-0.4[E(B-V) \times k_{1500}]},
\end{equation}

\noindent
is $f_{\rm esc, abs} (LyC) \approx 0.39$, with possible values between $\simeq 0.23-0.69$, allowed by the constraints on the IGM ($0.37 < T(IGM) < 1 $) and $f_{\rm esc, rel} (LyC)$ ($<1$). This a useful quantity to estimate the number of ionizing photons escaping from J0121$+$0025, $N_{\rm esc} \rm (LyC)$, such as:

\begin{equation}\label{photon}
    N_{\rm esc} (LyC) =  f_{\rm esc, abs} (LyC) \times N_{\rm int} (LyC),
\end{equation}

\noindent
where $N_{\rm int} \rm (LyC) \simeq 1.4 \times 10^{55}$~s$^{-1}$ is the intrinsic ionizing photon production rate from the S99 model scaled to $M_{\rm UV} = -24.1$ for a burst of 3~Myr and $Z_{\star}/Z_{\odot} = 0.4$ (and a \citealt{salpeter1955} IMF). We thus obtain an ionizing photon escape $N_{\rm esc} (LyC) \approx 6 \times 10^{54}$~s$^{-1}$ (with possible values ranging from $[3-10] \times 10^{54}$~s$^{-1}$).

\subsubsection{Possibility of foreground or AGN contamination}

We now discuss the possibility that the emission detected below $\lambda_{\rm obs} \simeq 3880$\AA{ }is due to foreground contamination leading to a false-positive detection of LyC signal of J0121$+$0025 or arising from an AGN.

\begin{figure}
  \centering
  \includegraphics[width=0.30\textwidth]{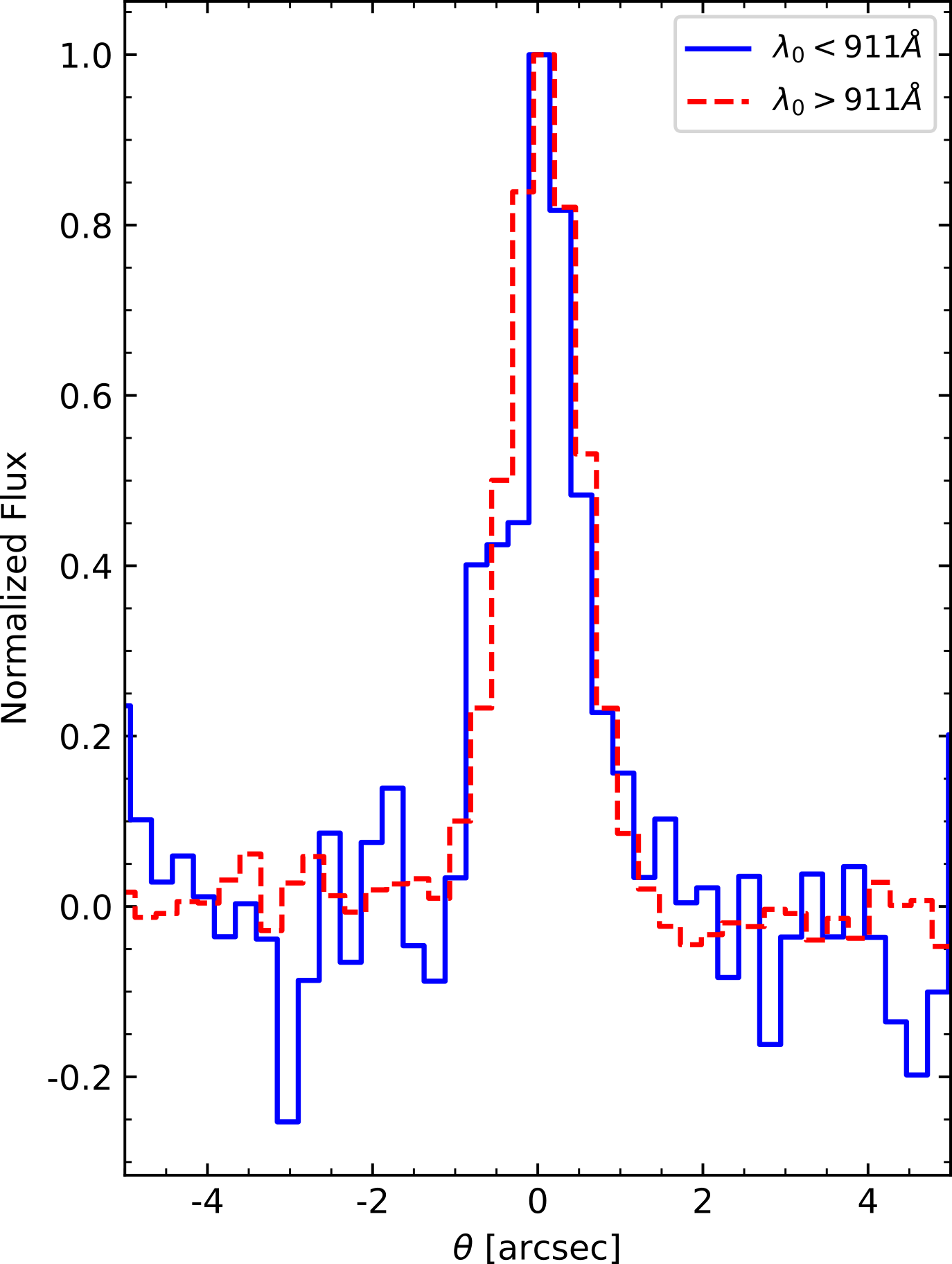}
  \caption{Comparison of the spatial profiles of the emission below (solid blue) and above (dashed red) the rest-frame $\lambda = 912$\AA{ }in the 2D spectrum of J0121$+$0025. These profiles have been extracted from the 2D spectrum over the rest-frame range $880-910$\AA{ }and $920-950$\AA, respectively, and both have consistent spatial morphologies and are co-spatial.}
  \label{fig6}
\end{figure}

In Figure \ref{fig6} the spatial profiles of the emission below and above $\lambda_{\rm 0} = 912$\AA{ }are compared. These profiles have been extracted from the 2D spectrum over the rest-frame range $880-910$\AA{ }and $920-950$\AA, respectively. Both profiles have consistent spatial morphologies with $\rm FWHM_{(\lambda_{\rm 0} < 912)} = 0.98^{\prime \prime} \pm 0.11^{\prime \prime}$ and $\rm FWHM_{(\lambda_{\rm 0} > 912)} = 1.02^{\prime \prime} \pm 0.02^{\prime \prime}$ which are compatible with being unresolved. Their centroids appear co-spatial, without any evidence for a spatial offset. 

From the best seeing condition image $R$-band ($\rm FWHM \simeq 0.5^{\prime \prime}$) we do not see any evidence for the presence of an additional source to J0121$+$0025, which is barely resolved only (Figure \ref{fig1}). Given the $5\sigma$ depth of $\simeq 26.5$ of this image, a possible contaminant would be easily detected if it was spatially offset from J0121$+$0025 by $\gtrsim 0.3^{\prime \prime}$, given the observed magnitude $m = 24.18$ measured from the spectrum at $\lambda_{\rm 0} < 912$\AA. 
In addition, one strong absorption line with a residual intensity compatible with zero is detected at $\lambda_{\rm obs} \simeq 4740$\AA, that is associated with Ly$\alpha$ from a H~{\sc i} absorption system at $z=2.898$ (\#2 in Figure \ref{fig5}). The non-detection of any flux at $\lambda_{\rm obs} \simeq 4740$\AA{ }below a $2\sigma$ level of $\simeq 1 \times 10^{-18}$~erg~s$^{-1}$~cm$^{-2}$~\AA$^{-1}$, where the OSIRIS spectrum is $\simeq 4$ times more sensitive than at $\lambda_{\rm obs} \simeq 3850$\AA, makes the presence of a contaminant very unlikely, because it requires an extremely low, and maybe unrealistic $\beta_{\rm UV} < -3.2$ ($2\sigma$) source to explain such color.

The presence of a relatively bright ($m = 24.18$), very compact ($r_{\rm eff} \lesssim 0.1^{\prime \prime}$), quasi co-spatial with J0121$+$0025 ($\lesssim 0.2^{\prime \prime}$), and very blue ($\beta_{\rm UV} < -3.2$) lower-$z$ interloper is still possible and cannot be completely ruled out, but it is highly unlikely.

We now discuss the possible contribution of an AGN to the LyC emission observed in J0121$+$0025. In Section \ref{sfg} we have shown that the contribution of an AGN to the UV luminosity of J0121$+$0025 should be small, if present. More specifically, the AGN contribution should be at least $\lesssim 25\%$, otherwise photospheric absorption lines, which are intrinsically very weak, would not be detected in the OSIRIS spectrum. Considering the most extreme case, i.e., a contribution of $25\%$ to the UV flux density, the AGN would have a rest-frame 1500\AA{ }flux density $f_{1500, \rm obs, AGN} \simeq 2.2\mu$Jy, corresponding to $M_{\rm UV} = -22.7$. 
Assuming that the observed $f_{900} \rm (obs) =0.781 \pm 0.099\mu$Jy arises from such AGN, we obtain $(f_{900} / f_{1500})_{\rm obs, AGN} \simeq 0.36$, which is a factor of $\simeq 3-7$ larger than the typical values observed in type-I AGNs with LyC detection at similar redshift and luminosities ($(f_{900} / f_{1500}) \rm (AGN) \simeq 0.05-0.14$; \citealt{steidel2001, micheva2017}) or in other bright QSOs \citep[e.g.,][]{cristiani2016}, possible corresponding to a nonphysical LyC escape fraction. Therefore, it is unlikely that the LyC emission arises from a type-I AGN. The presence of an obscured type-II AGN in J0121$+$0025 is more difficult to constrain, but its contribution to the LyC emission can be neglected, as these sources are by definition very obscured at short wavelengths.

An additional piece of evidence that the emission detected below $\lambda_{\rm obs} \simeq 3880$\AA{ }is related with escape of ionizing photons comes from the intrinsic properties of J0121$+$0025. In fact, J0121$+$0025 could be identified as a strong LyC leaker candidate, even excluding the direct information about the LyC detection. 
From the spectrum, J0121$+$0025 shows very weak LIS lines, both in terms of $EW_{0}$ and the residual intensity. 
It has been shown, from observations and simulations, that the residual intensity of LIS lines correlates with $f_{\rm esc} (LyC)$ \citep[e.g.][and Saldana-Lopez in prep.]{heckman2001, alexandroff2015, chisholm2018, mauerhofer2021}. Using the prescription given in \cite{chisholm2018}, the predicted $f_{\rm esc, abs}^{\rm pred} \rm (LyC) \approx 0.25$ from the residual intensity of the Si~{\sc ii} line (see Section \ref{s994}) is compatible, within the uncertainties, to that observed/inferred from the spectrum, $f_{\rm esc, abs} (\rm LyC) \approx 0.4$. In addition, the Ly$\alpha$ line shows a narrow profile with an intrinsic $\rm FWHM = 350 \pm 40$~km~s$^{-1}$ and with its peak closed to the systemic redshift, $v_{\rm peak} \simeq 120\pm50$~km~s$^{-1}$. The Ly$\alpha$ profiles in the confirmed LyC leakers analysed in \cite{steidel2018} and \cite{fletcher2019} have their peaks less redshifted than the LyC non-leakers (their Figure 26), suggesting low neutral gas column density where Ly$\alpha$ photons, and likely LyC, could escape more easily \citep{verhamme2015}. 
Other observational signatures shared by LyC leakers are also present in J0121$+$0025, such as the strong P-Cygni profiles and broad He~{\sc ii} emission \citep[e.g.,][]{vanzella2018, rivera2019, vanzella2020}, 
low dust attenuation ($E(B-V)\simeq 0.04$), compact morphology ($r_{\rm eff} \sim 1$~kpc) but large SFR (next section), and so large SFR surface density \citep[e.g.,][]{izotov2016}, or the evidence for strong outflows ($v_{\rm peak} \rm (LIS) \simeq -450$~km~s$^{-1}$).

\subsection{Multi-wavelength SED fitting}\label{s996}

\begin{figure}
  \centering
  \includegraphics[width=0.48\textwidth]{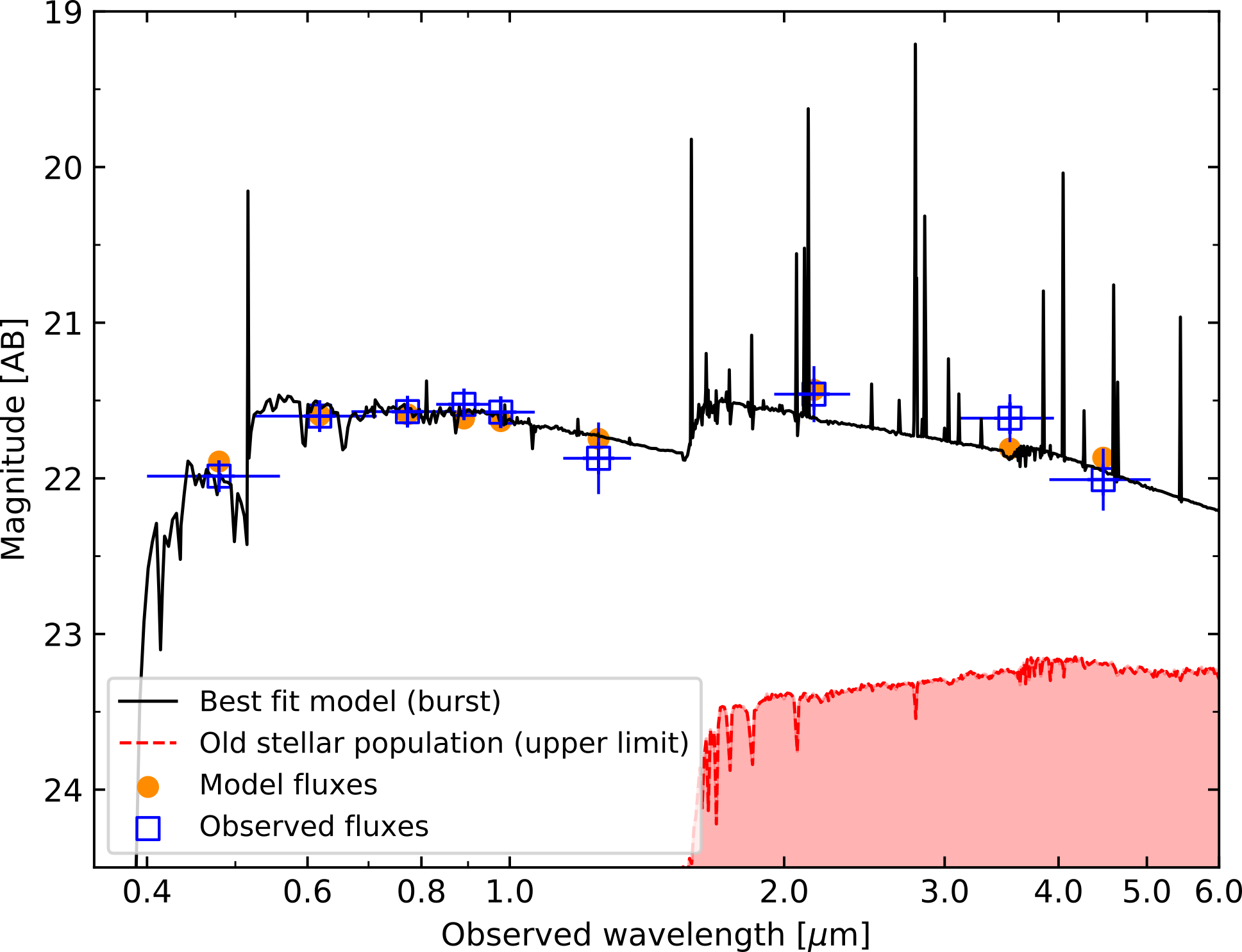}
  \caption{Best-fit model of the spectral energy distribution of J0121$+$0025 using CIGALE \citep{Burgarella2005}. The SED of J0121$+$0025 is dominated by a young and intense burst of star formation with SFR=981$\pm$232~$M_{\odot}$~yr$^{-1}$, an age of 7$\pm 2$~Myr, and a stellar mass of $\log (M_{\star}/M_{\odot}$)=9.9$\pm$0.1 (black line and orange circles). The fit uses photometry from $G$-band to IRAC $4.5 \mu$m (blue squares, see Table \ref{table1}). The red shaded region represents the upper limit SED of the old stellar component with an age of 500~Myr and $\log (M_{\star}^{\rm old}/M_{\odot}$)<10.4. }
  \label{fig7}
\end{figure}

Turning to the multi-wavelength properties, J0121$+$0025 shows a flat ($F_{\nu}$) spectral energy distribution (Figure \ref{fig7}) from optical to the mid-IR, that is broadly consistent with a young starburst. 

We perform SED-fitting with CIGALE code \citep{Burgarella2005, Boquien2019} using the photometry from $G$ to Spitzer $4.5\mu$m (Table \ref{table1}), covering a rest-frame wavelength $0.11 - 1.2\mu$m. 
The star-formation history (SFH) is modeled using two components: a young starburst with age $\leq 10$~Myr allowed by the analysis on the UV spectral features (see Section \ref{s991} and Figure \ref{fig2}), and a exponentially declining SFH with age of $500$~Myr. Using simultaneously two SFH components allows us to probe the properties of the young stellar population (i), that likely dominates the SED, and investigate the presence of an underlying old stellar component (ii). We adopted the stellar population models from \cite{bruzual2003}, and assume a fixed metallicity $Z/Z_{\odot}=0.4$ based on our analysis in Section~\ref{s991}. The \cite{calzetti2000} dust attenuation law is considered with $E(B-V)_{\star} < 0.1$, and assuming the ratio of the stellar and nebular $E(B-V)_{\star} / E(B-V)_{\rm neb} = 0.44$. We also adopt $f_{\rm esc, abs} \rm (LyC) \approx 0.40$.

Figure \ref{fig7} shows the best-fit model. The emission of J0121$+$0025 is dominated by a young stellar burst in the whole spectral range covered by the imaging data. The burst is characterized by a 10~Myr-weighted SFR=981$\pm$232~$M_{\odot}$~yr$^{-1}$, an age of 7$\pm 2$~Myr, and a stellar mass of $\log (M_{\star}/M_{\odot}$)=9.9$\pm$0.1. This yields a high specific SFR (sSFR=SFR/$M_{\star}$) of 98$\pm$32~Gyr$^{-1}$, which is a factor 20 times higher than $10^{10}M_{\odot}$ main-sequence star-forming galaxies at $3<z<4$ \citep[e.g.,][]{tomczak2016}. Table \ref{table2} summarizes the properties of J0121$+$0025. Both the age and the stellar mass of the burst are broadly consistent with the results inferred from the S99 best-fit model using the UV wind lines (age of $\sim 3$~Myr and $\log (M_{\star}/M_{\odot}$)=9.8, Section \ref{s991}).

\begin{table}
\begin{center}
\caption{Properties of J0121$+$0025. \label{table2}}
\begin{tabular}{l c c}
\hline \hline
\smallskip
\smallskip
  & Value & Uncertainty  \\
\hline 
R.A. (J2000)  & 01:21:56.09 & $0.1^{\prime \prime}$ \\
Dec. (J2000)  & $+$00:25:20.30 & $0.1^{\prime \prime}$ \\
$z_{\rm sys}$   & 3.244 & $0.001$   \\
$M_{\rm UV}$ (AB)   & $-24.11$ & $0.1$ \\
log(L[Ly$\alpha$/erg~s$^{-1}$]) & 43.8 & $0.1$  \\
$r_{\rm eff}$ (kpc)  & $<1.0$ & --- \\
Age (Myr)  & 3-7$^{a)}$   &   $<10^{a)}$  \\
$Z_{\star}/Z_{\odot}$     & 0.4   &   [0.2 - 1.0] \\
E(B-V)$_{\star}$  & 0.04   &   0.02   \\
SFR ($M_{\odot}$~yr$^{-1}$) & 981$^{b)}$ & 232$^{b)}$   \\
log($M_{\star}^{\rm burst}/M_{\odot}$)   & 9.9 & $0.1$ \\
sSFR (Gyr$^{-1}$)  & 98   & 32  \\ 
$\Sigma$SFR ($M_{\odot}$~yr$^{-1}$~kpc$^{-2}$) & $>157^{c)}$ & ---  \\
$f_{\nu} (900$\AA) ($\mu$Jy) & 0.781 & 0.099  \\
$f_{\nu} (900$\AA) / $f_{\nu} (1500$\AA) & 0.088 & 0.012   \\
$T(IGM)$ & 0.60 & 0.19  \\
$f_{\rm esc, rel}$ (LyC) & 0.56 & [0.34-1.0] \\
$f_{\rm esc, abs}$ (LyC) & 0.39 & [0.23-0.69]  \\
\hline 
\end{tabular}
\end{center}
\textbf{Notes. ---} (a) Age of the young stellar population obtained using UV wind lines ($\simeq 3$~Myr) and the best-fit model of the SED using CIGALE ($7 \pm 2$~Myr); (b) 10~Myr-weighted SFR obtained from the best-fit SED; (c) considering $r_{\rm eff} < 1$~kpc. 
\end{table}

On the other hand, the old stellar population is not well constrained. The best-fit gives a stellar mass for the old stellar component $\log (M_{\star}^{\rm old}/M_{\odot}$)=9.8$\pm$0.6. Figure \ref{fig7} shows the upper limit SED of the old stellar population (red), corresponding to $\log (M_{\star}^{\rm old}/M_{\odot}$)<10.4. Overall, the observed blue color around the rest-frame $\simeq 1.0\mu$m from the two IRAC/Spitzer channels $I1 - I2 = -0.40 \pm 0.31$ limits the presence of a $\log (M_{\star}^{\rm old}/M_{\odot}$)>10.4 old stellar component.

\section{Discussion}\label{discussion}

With $M_{\rm UV} = -24.1\pm0.1$ and $f_{\rm esc, abs} \rm (LyC) \approx 0.40$, J0121$+$0025 is not only one of the most UV-luminous star-forming galaxies ever discovered, but also the brightest LyC leaker known among star-forming galaxies. 
We now discuss the implications of such discovery.

\subsection{The brightest LyC emitter known}

J0121$+$0025 meets the two necessary conditions to be a strong LyC leaker: it is very efficient at producing hydrogen-ionizing photons and the properties of its ISM are favourable for their escape. 

The remarkably strong P-Cygni profiles in the wind lines O~{\sc vi}, N~{\sc v} and C~{\sc iv} seen in the spectrum of J0121$+$0025 (see Figures \ref{fig2} and \ref{fig3}) indicates unambiguously a very young age of the burst ($\simeq 3$~Mry) and the presence of a large number of O-type stars, the main-sequence stars hot enough to generate a significant number of ionizing photons. Strong P-Cygni profiles in these lines are ubiquitous in the spectra of LyC leakers, both at low-$z$ \citep[e.g.,][]{borthakur2014, izotov2016, izotov2018a, izotov2018b} and moderately high-$z$ \citep[e.g.,][]{rivera2019, vanzella2018, vanzella2020}, but appear weak or absent in composite spectra of more typical LBGs or LAEs \citep[e.g.,][]{shapley2003, du2018, nakajima2018, feltre2020, marques2020}. In contrast to the burstiness nature of J0121$+$0025, and likely other LyC leakers, smooth/continuous or declining star formation histories with ages from several tens to hundreds Myr could explain the weakness of these profiles in the spectra of typical LBGs/LAEs, which is in line with the SFHs usually inferred or assumed for them \citep[e.g.,][]{kornei2010, debarros2014, arrabal2020}. As a result, the amplitude and the age of the burst in J0121$+$0025 produces a large number of ionizing photons, $N_{\rm int} \rm (LyC) \simeq 1.4 \times 10^{55}$~s$^{-1}$ (see Section \ref{s995}), which is a factor of $\sim 10 - 30$ larger than that expected to be produced by a $M_{\rm UV}^{*} = -20.97$ galaxy \citep[e.g.,][]{reddy2009}, assuming a continuous star formation with 100~Myr age and the same metallicity.

\begin{figure*}
  \centering
  \includegraphics[width=0.98\textwidth]{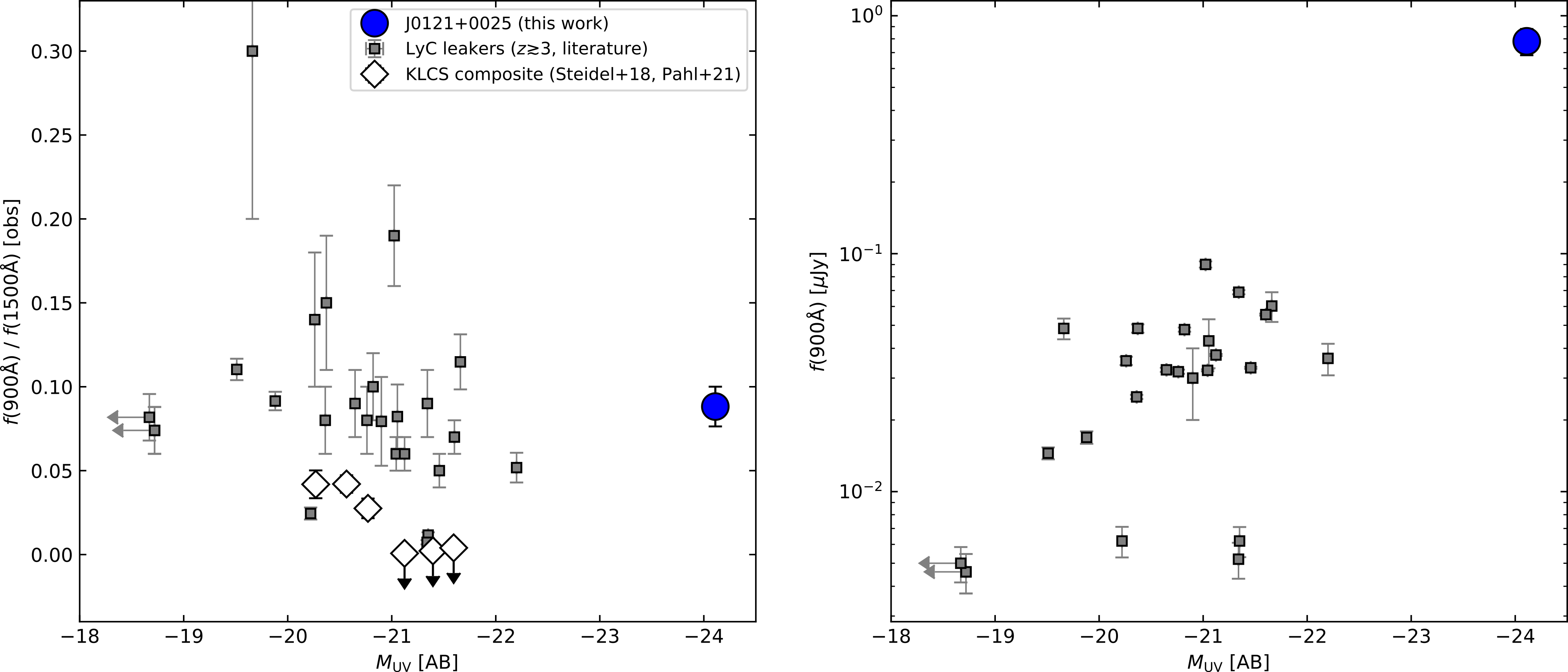}
  \caption{Relation between the observed ratio of the ionizing to non-ionizing flux density (in $F_{\nu}$ units, left) and the observed flux density at $\simeq 900$\AA{ }(in $\mu$Jy, right) versus the UV absolute magnitude. J0121$+$0025 is represented with a blue circle. For comparison, we also show the other $z \gtrsim 3$ star-forming galaxies known with significant detection of LyC radiation \citep[grey squares;][]{debarros2016, shapley2016, vanzella2016, steidel2018, vanzella2018, fletcher2019, ji2020}. Empty diamonds mark the position of the estimates of several composites from \protect\cite{pahl2021} in different bins of $M_{\rm UV}$. }
  \label{fig8}
\end{figure*}

The presence of strong P-Cygni profiles could be a potential indicator of LyC leakage, as discussed in previous works \citep[e.g.,][]{izotov2018b, chisholm2019}, at least for moderately metal-rich galaxies \citep[because the strength of the P-Cygni is also metallicity dependent, see:][]{izotov2021}. Note, however, that the presence of such wind lines indicates a high production efficiency of LyC photons, not necessarily their escape. Nevertheless, feedback from the strong winds of massive stars together with SN explosions expected in the early phase of a burst could play a major role in shaping the ISM, creating cavities of ionized gas where ionizing photons could escape more efficiently \citep[e.g.,][]{heckman2011, trebitsch2017}. This might be the case of J0121$+$0025.
The particular conditions of the ISM in J0121$+$0025 are in fact favorable for the escape of LyC radiation. LyC photons are easily absorbed by dust and neutral gas and these sources of opacity are apparently weak in J0121$+$0025, at least from our line-of-sight. The inferred $\beta_{\rm UV} = -2.05$ from the spectrum is compatible with low attenuation by dust, corresponding to $E(B-V)=0.04$. In addition, ISM absorption lines are very weak, with $EW_{0} \lesssim 1$\AA, and show residual intensities of $I/I_{0} \simeq 0.8$ for Si~{\sc ii}~1260\AA{ }and C~{\sc ii}~1334\AA, even though they are likely saturated (Section \ref{s994}). These findings suggest a clumpy ISM with a non-unity covering fraction ($C_{f}$ (H~{\sc i}) $\simeq 0.55$, \citealt{gazagnes2018}). The detection of blueshifted profiles in LIS lines in J0121$+$0025 ($\simeq -450$~km~s$^{-1}$, see Section \ref{s994} and Figure \ref{fig3}) supports the presence of such strong outflows.

We now compare in Figure \ref{fig8} the LyC properties of J0121+0025 with those from other confirmed LyC leakers at $z \gtrsim 3$. The comparison sample consists of $\sim 20$ sources with significant detection of LyC radiation taken from \cite{debarros2016}, \cite{shapley2016}, \cite{vanzella2016}, \cite{steidel2018}, \cite{vanzella2018}, \cite{fletcher2019} and \cite{ji2020}. The left panel compares the observed ratio of the ionizing to non-ionizing flux density, $(f_{900} / f_{1500})_{\rm obs}$, versus the UV absolute magnitude.  We use $(f_{900} / f_{1500})_{\rm obs}$ because it is model-independent and does not rely on assumptions about the properties of the underlying stellar population, such as SFHs, age and metallicity. Furthermore we do not correct these values for the IGM absorption, because it is highly uncertainty and it is not always available. 
From this figure, our results indicate that a significant fraction of LyC photons can escape in sources with a wide range of UV luminosity, from UV faint ($M_{\rm UV} \gtrsim -18.7$,  \citealt{fletcher2019}) to extremely UV luminous ($M_{\rm UV} = -24.1$, this work), and, therefore, are not restricted to the faintest ones as previously thought (\citealt{steidel2018, pahl2021}, see also: \citealt{bian2020}).

The importance of detecting LyC emission in such a luminous source like J0121+0025 is highlighted in the right panel of Figure \ref{fig8}. Here, we compare the observed flux density at $\simeq 900$\AA, $f_{900}$. Again, no IGM corrections have been applied nor corrections for the luminosity distance, although all but two LyC leakers \citep{vanzella2018, ji2020} are roughly at the same redshift of J0121+0025. Therefore, this comparison should be treated with care, serving for illustrative purposes only. 
The combination of the extreme luminosity of the starburst ($M_{\rm UV} = -24.1$) and the large $f_{\rm esc}$~(LyC) makes J0121$+$0025 the brightest and the most powerful LyC emitter known among the star-forming galaxy population. The observed LyC flux in J0121$+$0025 is in fact comparable to the sum of the LyC flux of all star-forming galaxies with LyC leakage known at these redshifts. Only the highly magnified \textit{Sunburst Arc} ($\mu 
\sim 30-100 \times$, \citealt{pignataro2021}) or bright QSOs have comparable observed LyC fluxes ($\approx 1 \mu$Jy, \citealt{steidel2001, lusso2015, rivera2019}).

The discovery of such a powerful LyC emitter raises now the question regarding the role of UV-luminous star-forming galaxies to the cosmic reionization (e.g., \citealt{sharma2016}, \citealt{Naidu2020}). Answering this question is however out of the scope of this work, as it requires the knowledge of two fundamental properties that are highly uncertain: the volume density of such luminous sources at $z\gtrsim 7$ and the physical properties connecting LyC leakage. 
Nevertheless, we can place rough constraints on the UV ionizing background at $z\sim 3$. To do so, we assume that luminous sources ($M_{\rm UV} < -22$) share the same properties as J0121$+$0025, i.e., $f_{\rm esc, abs} \rm (LyC) \sim 0.40$ and  log($\xi_{\rm ion} ) = 25.2$. The co-moving production rate of hydrogen-ionizing photons ($\dot{N}_{\rm ion}$ ) is given by: 

\begin{equation}\label{photon}
    \dot{N}_{\rm ion} =  f_{\rm esc, abs} (LyC) \: \xi_{\rm ion} \: \rho_{\rm UV} \; \rm [s^{-1} Mpc^{-3}],
\end{equation}

\noindent
where $\rho_{\rm UV}$ is the dust-corrected UV luminosity density. For $\rho_{\rm UV}$, we integrate the UV luminosity function of \cite{reddy2009} down to $-22$~AB (or $\gtrsim 3 L_{\rm UV}^{*}$) and assume $E(B-V)=0.04$. This yields a $\dot{N}_{\rm ion} \sim 10^{49.8}$~s$^{-1}$~Mpc$^{-3}$, which is fairly lower than that provided by  QSOs at $z\sim 3$ ($\dot{N}_{\rm ion} \rm (QSO) \sim 10^{50.5} - 10^{51.0}$~s$^{-1}$~Mpc$^{-3}$, e.g., \citealt{becker2013}, \citealt{cristiani2016}). 
However, the situation may differ at very high-$z$. Recent studies have found remarkably bright galaxies at $z\gtrsim 7$ that are in excess compared to the generally observed Schechter function of luminosity functions \citep[e.g.,][]{bowler2014, oesch2014, bowler2015, ono2018, stefanon2019}. Assuming the double
power-law luminosity function at $z\sim 6$ of \cite{bowler2015}, we find $\dot{N}_{\rm ion} \sim 10^{48.9}$~s$^{-1}$~Mpc$^{-3}$ for star-forming galaxies brighter than $-22$~AB, that is comparable to that inferred in QSOs at these redshifts ($\dot{N}_{\rm ion} \rm (QSO) \sim 10^{48.8}$~s$^{-1}$~Mpc$^{-3}$;  \citealt{matsuoka2018c}).
On the other hand, it is not clear if the properties connecting the LyC leakage in J0121$+$0025, and therefore the large $f_{\rm esc}$~(LyC), can be applied to other bright star-forming galaxies \citep[see e.g.,][]{harikane2020}. This will be discussed in the next section.

\subsection{Understanding UV-luminous star-forming galaxies: diverse properties and insights for LyC leakage}

J0121$+$0025 shows intriguing properties that differ from those expected in bright star-forming galaxies.  
Here we discuss some of these properties and compare them with other UV-luminous star-forming galaxies, with particular emphasis to the properties that could be related with the LyC leakage. 

A well established trend relating $M_{\rm UV}$ and the strength of Ly$\alpha$ and ISM lines has been found in previous works \citep[e.g.,][]{shapley2003, vanzella2009, trainor2015, du2018}, for which Ly$\alpha$ is found to be weak and LIS lines strong in UV-bright sources. While this trend has been established with statistical significance for galaxies with $M_{\rm UV}$ between $-18$ and $-21$, a few other known $M_{\rm UV} \simeq -23$ LBGs (\citealt{des2010, lee2013b, marques2018, harikane2020}) show the same trend, presenting a completely damped Ly$\alpha$ absorption and strong LIS lines ($EW_{0} \simeq 2 -4$\AA) in their spectra, which is compatible with a large column density of neutral gas ($N$(H~{\sc i}$)>10^{20}$~cm$^{-2}$). However, such trend is not seen in J0121$+$0025 as it shows weak LIS lines and a relatively strong Ly$\alpha$ line. Another interesting difference is that P-Cygni in wind lines, in particular in N~{\sc v} that is very sensitive to the age, are weak or absent in these $M_{\rm UV} \simeq -23$ LBGs \citep[e.g.][]{des2010, marques2018, harikane2020}, which could indicate that their UV continua is not dominated by O-type stars ($\leq 10$~Myr), at least compared to J0121$+$0025, but by older and less luminous stars (e.g., B-type stars).
In addition, vigorous star-forming galaxies are also found to be more dusty, because the production of dust is tightly linked with star formation. However, this does not apply for J0121$+$0025. In fact, the derived $ \rm SFR = 981 \pm 232$~$M_{\odot}$~yr$^{-1}$ of J0121$+$0025 is comparable, in absolute terms, to those inferred in very dusty, far-IR bright systems, but as opposed to them, only a small fraction of the total SFR of J0121$+$0025 is obscured ($\simeq 30\%$). 

On the other hand, the properties of J0121$+$0025 are remarkably similar to those observed in a few, very young, and also luminous starbursts, like the recently discovered extremely luminous starburst galaxy BOSS-EUVLG1 at $z=2.469$ ($M_{\rm UV} \simeq -24.4$; \citealt{marques2020b}), or the less, but still luminous LyC leaker \textit{Ion3} galaxy at $z\simeq 4.0$ ($M_{\rm UV} \simeq -22.2$; \citealt{vanzella2018}). 
The spectra of these galaxies show very strong P-Cygni profiles in wind lines (O~{\sc vi}, N~{\sc v} and C~{\sc iv}), intense nebular emission in rest-frame optical lines ($EW_{0} (\rm H\alpha) \gtrsim 700$\AA) and SEDs that are compatible with a very young and intense starburst of a few Myr ($\lesssim 10$~Myr). As J0121$+$0025, these galaxies show very weak ISM absorption lines ($EW_{0} < 1$\AA), strong Ly$\alpha$ emission $EW_{0} > 20$\AA{ }and are almost un-obscured ($\beta_{\rm UV} < -2.2$; \citealt{vanzella2018, marques2020b}). Other interesting properties shared in these galaxies are the high specific SFR ($\rm sSFR \simeq 90-100$~Gyr$^{-1}$), compact morphologies ($r_{\rm eff} \sim 1$~kpc or less), and therefore high star-formation rate surface density ($\Sigma \rm SFR \gtrsim 100$~$M_{\odot}$~yr$^{-1}$~kpc$^{-2}$), properties that are common in other LyC leakers \citep[e.g.,][]{izotov2016, izotov2018b} and are thought to play a key role in transforming the ISM structure (feedback). 

Understanding such diversity in the properties of UV-luminous galaxies is challenging, and possibly premature for now given the lack of statistics. We note that only a handful of luminous sources are known as bright as $M_{\rm UV} \sim -23$ and only two brighter than $M_{\rm UV} \sim -24$, J0121$+$0025 (this work) and BOSS-EUVLG1 \citep{marques2020b}. 
Nevertheless, the differences are already notorious, in particular those thought to be closely related to the LyC leakage. 

A possible explanation to these differences may be simply related to different stages in the evolution of galaxies. J0121$+$0025, BOSS-EUVLG1 or \textit{Ion3} could represent a vigorous starburst seen at very initial stages ($<10$~Myr), when the starlight is dominated by young stars and before dust-attenuation become efficient, enhancing the UV luminosity.
The extreme feedback expected in such early, intense, and likely short-lived phase could eject the gas and dust from star-forming regions, allowing the escape of LyC photons  \citep[e.g.,][]{sharma2017, trebitsch2017, arata2019}. 
In fact, a powerful ionized gas outflow has been detected in BOSS-EUVLG1 with a log($M_{\rm out}/M_{\odot}) \simeq 8$ and an outflowing velocity $v_{\rm out} \simeq 600$~km~s$^{-1}$ \citep{alvarez2021}. 
On the other hand, if the SFR drops significantly at later stages, the UV luminosity will be still high for a considerable period of time ($\sim 50$~Myr) due to the contribution of B-type stars, but in such case SN and stellar feedback could be less effective in clearing sight lines, and neutral gas and dust would cover star-forming regions absorbing the LyC radiation. In both cases, the galaxy will appear bright in the UV, but some spectral features would appear dramatically different. 
Alternatively, the differences seen in the spectra of UV-luminous galaxies could arise by a non-homogeneous distribution of gas and dust in these UV-luminous galaxies, for which some of them would have star-forming regions cleared by dust and gas from a favourable sight-line (J0121$+$0025, BOSS-EUVLG1 or \textit{Ion3}), while others not \citep[e.g.,][]{harikane2020}. 

Independently of the scenario invoked to explain these differences, the discovery of J0121$+$0025 (and \textit{Ion3}, \citealt{vanzella2018}) indicates that, at least some UV-luminous star-forming galaxies can be strong LyC emitters, which contradicts recent findings from \cite{harikane2020}. We note that the sample of $M_{\rm UV} \simeq -23$ LBGs at $z\sim 6$ analysed in \cite{harikane2020} was previously selected to have log(L[Ly$\alpha$/erg~s$^{-1}]) < 43.0$ \citep{matsuoka2016, matsuoka2018a, matsuoka2018b, matsuoka2019} to avoid a possible contamination of an AGN. However, such threshold is conservative \citep[see e.g.,][for other luminous LAEs]{ouchi2009, sobral2015, matthee2017} and can yield to a selection bias towards high $N$(H~{\sc i}) and/or declining SFHs. In fact, the composite spectrum presented in \cite{harikane2020} shows a completely damped Ly$\alpha$ absorption, which is compatible with a large column density of neutral gas, and naturally explain the strong LIS lines observed in the spectrum. On the other hand, other UV-bright sources identified as AGNs by \cite{matsuoka2016, matsuoka2018a, matsuoka2018b, matsuoka2019}, based only on log(L[Ly$\alpha$/erg~s$^{-1}]) > 43.0$, show intense and narrow Ly$\alpha$ emission (as narrow as $<230$~km~s$^{-1}$), very weak LIS lines, and, interestingly, evidence of strong P-Cygni in N~{\sc v} (see Figure 9 in \citealt{matsuoka2019}). The authors argue that such a P-Cygni profile could arise from a weak BAL-QSO. However, as shown here in this work, similar profiles could be naturally explained by a very young and hot stellar population that could enhance the UV and Ly$\alpha$ luminosities in these sources, similar to what is happening in J0121$+$0025 and BOSS-EUVLG1 \citep{marques2020b}.

In closing, it is clear that properties of UV-luminous star-forming galaxies are still not very well understood and must be investigated in more detail with a large statistical sample. In a future work, we will present a large sample of other $\sim 70$ extremely UV-luminous star-forming galaxies discovered within the eBOSS survey (R. Marques-Chaves, in prep.), hoping to answer some of these important questions.

\section{Summary and Conclusion}\label{conclusion}

This work reports the discovery of J0121$+$0025 at $z=3.244 \pm 0.001$, an extremely luminous in the UV ($M_{\rm UV} \simeq -24.11$, AB) and Ly$\alpha$ line (log[$L_{\rm Ly \alpha} / \rm erg~s^{-1}] = 43.8$) star-forming galaxy with copious emission in the LyC spectral range ($\lambda_{0} < 912$\AA). 
J0121$+$0025 is a compact starburst, with $r_{\rm eff} = 1\pm0.5$~kpc, that is only barely resolved in very good seeing conditions ground-based imaging. The optical to mid-IR photometry is dominated by the emission of a vigorous starburst, with log($M_{\star}^{\rm burst}/M_{\odot}) = 9.9\pm0.1$ and a 10~Myr-weighted $\rm SFR = 981 \pm 232$~$M_{\odot}$~yr$^{-1}$. This yields a high specific star-formation rate $\rm sSFR = 98 \pm 32$~Gyr$^{-1}$ and a SFR density $\Sigma \rm SFR > 157$~$M_{\odot}$~yr$^{-1}$~kpc$^{-2}$ (considering $r_{\rm eff} < 1$~kpc).
The high SNR OSIRIS/GTC spectrum of J0121$+$0025 reveals strong P-Cygni in wind lines of O~{\sc vi}, N~{\sc v} and C~{\sc iv}, which are well reproduced by a starburst model with an extremely young age of $\simeq 3$~Myr, which is roughly consistent with the age derived from the multi-wavelength SED ($7\pm2$~Myr). The spectrum shows a rest-frame UV slope $\beta_{\rm UV} = -2.05 \pm 0.10$, consistent with low dust attenuation $E(B-V)_{\star} = 0.04 \pm 0.02$. 
It also shows other features characteristic of star-forming galaxies, such as stellar absorption originated in the photospheres of hot stars, for which a significant contribution of an AGN to the luminosity is ruled out. The Ly$\alpha$ is moderately strong ($EW_{0} \rm [Ly\alpha] = 14 \pm 3$\AA) and shows a narrow profile ($\rm FWHM \simeq 350$~km~s$^{-1}$) with its peak redshifted, but close to the systemic velocity, by $\simeq 120$~km~s$^{-1}$. Low-ionization ISM lines are also detected, but appear much weaker when compared to those observed in typical LBGs. Both the weakness ($EW_{0} \rm [LIS] \simeq 1$\AA) and the large residual intensity ($I/I_{0}) \simeq 0.8$) suggest a clumpy geometry with a non-unity covering fraction or a highly ionized ISM, for which a significant fraction of ionizing photons could escape. LyC radiation is detected with a significance of $\simeq 7.9$ in the OSIRIS spectrum, corresponding to a flux density $f_{900} = 0.781 \pm 0.099 \mu$Jy and an ionizing to non-ionizing flux density $(f_{900} / f_{1500})_{\rm obs} = 0.09 \pm 0.01$. The contribution of a foreground or AGN contamination to the LyC signal is discussed in detail, and although it cannot be completely ruled out, it is very unlikely. 
This makes J0121$+$0025 the most powerful LyC emitter known among the star-forming galaxy population.
Our results indicate that at least some UV-luminous star-forming galaxies are strong LyC leakers, bringing new insights to the discussion of the role of luminous and very young starbursts to the cosmic reionization.

\section*{Acknowledgements}

The authors thank the referee, Eros Vanzella, for useful comments that greatly improved the clarity of this work.
Based on observations made with the Gran Telescopio Canarias (GTC)  installed in the Spanish Observatorio del Roque de los Muchachos of the Instituto de Astrof\'{i}sica de Canarias, in the island of La Palma. 
J.A.M., L.C., and I.P.F. acknowledge support from the Spanish State Research Agency (AEI) under grant numbers ESP2015-65597-C4-4-R, ESP2017-86852-C4-2-R, RyC-2015-18078,  PGC2018-094975-B-C22, and MDM-2017-0737 Unidad de Excelencia ''Mar\'{i}a de Maeztu''- Centro de Astrobiolog\'{i}a (CSIC-INTA). A.S.L. acknowledge support from Swiss National Science Foundation.

\section*{Data availability}
The data underlying this article will be shared on reasonable request to the corresponding author.

\bibliographystyle{mnras}
\input{main.bbl}

\label{lastpage}
\end{document}